%% file: acl_latex.tex
\documentclass[11pt]{article}

\usepackage[preprint]{acl}

\usepackage{times}
\usepackage{latexsym}
\usepackage[T1]{fontenc}
\usepackage[utf8]{inputenc}
\usepackage{microtype}
\usepackage{inconsolata}
\usepackage{graphicx}
\usepackage{url}
\usepackage{amsmath}
\usepackage{amssymb}
\usepackage{booktabs}
\usepackage{tabularx}
\usepackage{array}
\usepackage{multirow}
\usepackage{xspace}

\newcommand{\method}{MaskClaw\xspace}
\newcommand{\dataset}{P-GUI-Evo\xspace}

\title{\method: Edge-Side Personalized Privacy Arbitration for GUI Agents with Behavior-Driven Skill Evolution}

\author{
Yanqiu Zhao\textsuperscript{1,\textdagger} \quad
Dongying Zheng\textsuperscript{1,\textdagger} \quad
Kaibo Huang\textsuperscript{1,*} \quad
Yukun Wei\textsuperscript{1} \quad
Zhongliang Yang\textsuperscript{1,*} \quad
Linna Zhou\textsuperscript{1} \\
\textsuperscript{1}Beijing University of Posts and Telecommunications \\
\textsuperscript{\textdagger}Equal contribution. Authors are listed in alphabetical order. \quad
\textsuperscript{*}Corresponding authors. \\
\texttt{\{ZYQiu, zhengdongying, Huangkaibo, weiyukun, yangzl, zhoulinnna\}@bupt.edu.cn}
}

\begin{document}
\maketitle

\begin{abstract}
GUI agents rely on screenshots to infer intent and operate across applications, but these screenshots often contain private messages, medical records, payment credentials, and workplace-specific workflows. Privacy decisions in this setting depend on task, recipient, application state, and user role, yet static PII detectors miss these boundaries and cloud-side VLM reasoning can upload the raw screen before deciding what should be protected. We present \method, an edge-side privacy arbitrator for GUI agents. \method{} extracts local visual evidence, retrieves user- and task-specific policy memory, and decides Allow, Mask, or Ask before raw screenshots leave a trusted user- or organization-controlled environment. In five designed skill-evolution scenarios, it turns corrections, cancellations, and edits into reusable privacy skills checked by a sandbox gate. We introduce \dataset, a benchmark built from real UI patterns, reconstructed HTML screens, and sanitized labels. Experiments show that pattern matching, cloud reasoning, and routing alone tend to over-confirm, over-mask, or expose raw screenshots under the same protocol. The artifact is available at \url{https://github.com/Theodora-Y/MaskClaw}.. 
\end{abstract}

\section{Introduction}

Graphical user interface (GUI) agents automate mobile and desktop tasks by interpreting screenshots and acting on user instructions \citep{zhang2025,liu2024,wang2024}. Multimodal foundation models and GUI grounding methods drive this progress \citep{hong2024,zheng2024,wang2025}, while compact multimodal models make edge-side arbitration feasible for extracting local evidence and deciding what observation may leave the device \citep{hu2024,yao2024}. The edge side may be the endpoint itself or a trusted local environment where the AI model, large-language-model component, or privacy computation runs, including phones, PCs, in-vehicle systems, smart-home devices, workstations, personal servers, and private enterprise servers.

The same visual interface creates a privacy bottleneck. Screenshots may contain private messages, account identifiers, medical records, financial tables, business procedures, or task-irrelevant personal details. Prior work shows that language and vision-language models can leak, infer, or be induced to reveal sensitive information from textual and visual inputs \citep{carlini2021,kim2023,tomekce2024,wang2026,zhang2026}. Sending raw screenshots to a remote VLM creates a cloud-dependency paradox: the screen is exposed before protection is decided.

\begin{figure}[t]
\centering
\includegraphics[width=\linewidth,trim=160 250 195 0,clip]{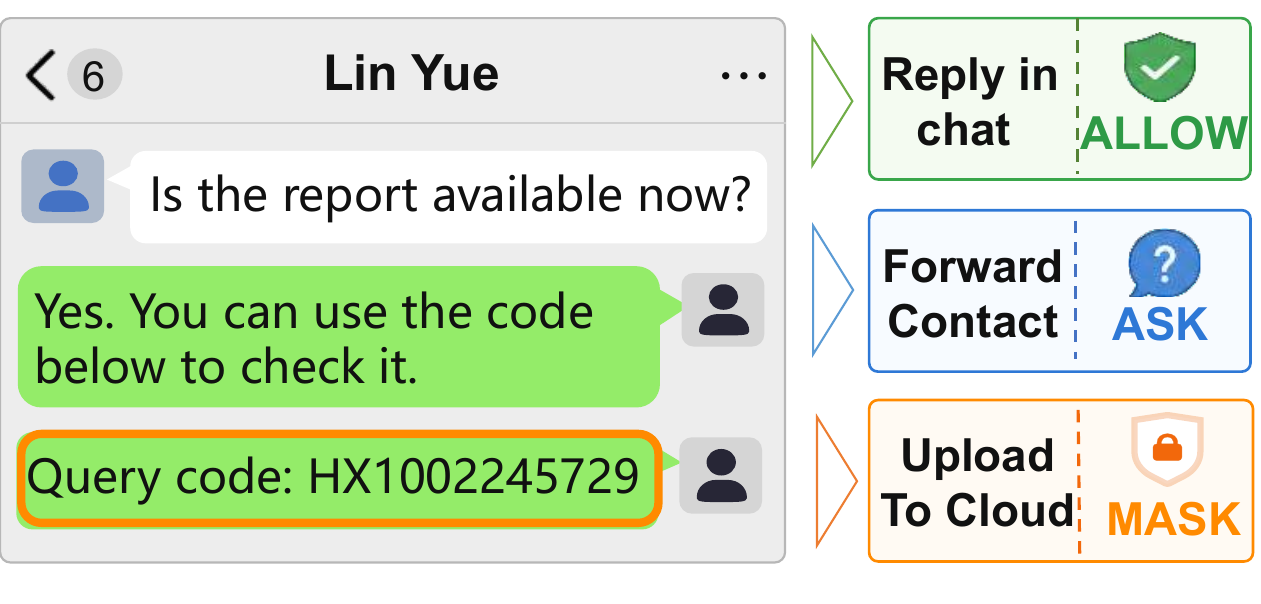}
\caption{Overview of the motivation and problem setting. GUI agents need screenshots to act, but the same screenshots can expose task-irrelevant private content. \method{} treats privacy as edge-side arbitration: decide Allow, Mask, or Ask before raw screenshots leave the trusted environment.}
\label{fig:introduction}
\end{figure}

Existing privacy mechanisms leave this decision boundary unresolved. Pattern-based tools detect common PII \citep{habernal2023}, and prompt-side masking can reduce online LLM exposure \citep{zhu2024}. GUI-agent privacy depends on context and personal policy. A value that is routine in a reimbursement form may require protection when it appears as a bank balance, and the same screen can be released for one persona and require confirmation for another. The correct action depends on screen content, task intent, recipient, application state, and user role \citep{lu2024,chaudhari2026,wang2025a,salemi2023}. We arbitrate exposure on the trusted edge before releasing raw screens, with policies refined from user behavior.

In summary, our main contributions are:
\begin{itemize}
     \item We propose \method, an edge-side privacy arbitrator that uses local evidence and policy memory to choose Allow, Mask, or Ask before raw screenshots leave a trusted environment.
    \item We introduce a feedback-to-skill loop that turns corrections, cancellations, edits, and negative feedback into reusable privacy skills checked by rule tests, held-out cases, and sandbox validation.
    \item We construct \dataset, a sanitized, HTML-reconstructed benchmark for multi-persona GUI privacy, covering local policy memory, rule-free routes, cloud and routing controls, and skill evolution under one protocol.
\end{itemize}

\section{Related Work}

\subsection{GUI grounding for agentic screen control}
GUI agents such as AppAgent, AutoGLM, Mobile-Agent, CogAgent, and Step-GUI combine visual state understanding with action execution \citep{zhang2025,liu2024,wang2024,hong2024,yan2025}. GUI grounding methods and benchmarks structure screenshots into actionable observations and evaluate device-control behavior \citep{yang2023,lu2024b,you2024,zheng2024,rawles2023,xu2025}. \method{} builds on these perception and control capabilities to govern which visual evidence may be released to a downstream or cloud agent.

\subsection{Contextual privacy for visual agent observations}
Closest to our setting, VLM and GUI-agent privacy benchmarks show that visual inputs can leak private or socially contextual information \citep{lu2024,chaudhari2026,tomekce2024,wang2026,zhang2026,wang2026a}. GUI anonymization uses type-preserving placeholders to keep interfaces usable while hiding visual content \citep{zhao2026}; MemPrivacy protects edge-cloud agent memory through local privacy-span extraction and type-aware placeholder substitution \citep{chen2026}. Agentic settings add further risk because prompt injection and privacy jailbreaks can turn one exposed observation into downstream leakage \citep{chang2026,wang2025b}. These methods reduce exposure while leaving the live Allow/Mask/Ask decision open under a specific task, recipient, application state, and user role. \method{} treats this choice as context-conditioned policy arbitration, consistent with contextual integrity \citep{nissenbaum2004a}, before raw screenshots leave the trusted edge boundary.

\subsection{Feedback-driven optimization of agent policies}
Language feedback can optimize prompts, reasoning traces, and agent behavior through textual search, as shown by OPRO, TextGrad, Reflexion, STaR, evolutionary prompt optimization, and WebRL \citep{yang2024,yuksekgonul2024a,shinn2023,zelikman2022,guo2024,qi2025}. Recent GUI skill systems explore verifiable tree evolution, recursive skill-RL, collective skill evolution, and memory-designed reusable skills \citep{jiang2026,xia2026,ma2026,zhou2026}. \method{} re-purposes this paradigm for personalized privacy skills, keeping candidate updates only after sandbox validation confirms policy alignment and safe exposure control.

\method{} connects GUI grounding, privacy protection, and skill optimization through local evidence, context-conditioned arbitration, and auditable privacy-skill updates.

\section{Problem Formulation and Benchmark}

\subsection{Personalized Privacy Decision}

We formulate GUI privacy arbitration as a personalized, context-conditioned classification problem. The arbitrator decides whether visual evidence may be exposed under a specific user, task, recipient, and application context. The same screen element can receive different decisions across personas or workflows because privacy depends on user-specific context beyond input content alone \citep{salemi2023,zhang2026}.

For the $i$-th GUI-agent interaction, we define the prediction instance as:
\begin{equation}
x_i=(I_i,q_i,a_i,c_i,p_i),
\end{equation}
where $I_i$ is the current GUI screenshot, $q_i$ the user instruction or task goal, $a_i$ the candidate intent or next action of the GUI agent, $c_i$ the application platform and relational context, and $p_i$ the persona and policy context. An arbitrator predicts $y_i=f(x_i)$ with output space $\mathcal{Y}=\{\mathrm{Allow},\mathrm{Mask},\mathrm{Ask}\}$, corresponding to no intervention, required protection, and required user consent.

\subsection{The \dataset{} Benchmark}

To evaluate personalized privacy arbitration and behavior-driven skill evolution, we construct \dataset, a real-UI-pattern-derived, HTML-reconstructed benchmark of 832 GUI privacy samples. The benchmark covers 296 normalized scenario cores, with 2.81 samples per core on average. \dataset{} connects GUI-agent and mobile-control benchmarks, which focus on interaction and task completion, with VLM and GUI-agent privacy benchmarks, which focus on visual leakage or privacy assessment \citep{rawles2023,xu2025,lu2024,wang2026,zhang2026,wang2026a}. It evaluates Allow/Mask/Ask decisions from GUI observations and task context, including cases where similar visual evidence requires different decisions across personas or workflows.

We build each record through a controlled reconstruction and audit workflow. We start from common GUI interaction patterns and platform-like layouts, using placeholders, fake identities, or empty form states as sanitized surrogates for private content. An LLM drafts the HTML structure and scenario text and supports PII-category consistency checks. The project authors fix task content and audit plausibility, fluency, alignment with the candidate agent intent, and support for the expected decision. OCR checks and PII audits verify that visible sensitive evidence can be detected and that the rendered screenshot matches the scenario metadata.

Each sample contains a GUI screenshot, application platform, persona context, relationship context, user instruction, agent intent, benchmark bucket, and an expected Allow/Mask/Ask decision for offline evaluation. The final joint-label distribution is Mask=438, Allow=314, and Ask=80. Prediction methods receive only runtime-available fields; risk tags, PII categories, and lower-level audit labels are reserved for analysis. Scenario cores are expanded into structural, visual, or lexical variants that preserve the privacy boundary. Among the 296 cores, 144 have at least one expanded variant, with 173 lexical, 201 structural, and 208 visual-shift rows.

\begin{table}[t]
\centering
\small
\begin{tabular*}{\linewidth}{@{}p{.29\linewidth}@{\extracolsep{\fill}}r@{\hspace{0.4em}}p{.53\linewidth}@{}}
\toprule
\textbf{Partition} & \textbf{Count} & \textbf{Purpose}\\
\midrule
User A & 234 & healthcare and medical workflow\\
User B & 246 & commerce and livestream selling\\
User C & 352 & office and everyday services\\
\midrule
D1 basic & 174 & clean in-distribution cases\\
D2 generalization & 546 & UI/task variants\\
D3 stress & 112 & noisy boundary cases\\
\bottomrule
\end{tabular*}
\caption{Persona and bucket distribution in \dataset.}
\label{tab:dataset}
\end{table}

Table~\ref{tab:dataset} summarizes the persona and bucket distribution. User A/B/C are synthetic persona identifiers for varying workflow, recipient, and task context. D3 stress cases include OCR noise, occlusion, low resolution, mixed languages, popup interference, or multiple sensitive regions.

\begin{table}[t]
\centering
\small
\begin{tabular*}{\linewidth}{@{}p{.63\linewidth}@{\extracolsep{\fill}}r@{}}
\toprule
\textbf{Construction or audit statistic} & \textbf{Count}\\
\midrule
Active benchmark samples & 832\\
Normalized scenario cores & 296\\
Cores with expanded variants & 144\\
Samples with non-\texttt{None} PII type & 715\\
Location-ready PII evidence samples & 685\\
Localizable evidence boxes & 1,529\\
L1/L3 usable samples & 736\\
\bottomrule
\end{tabular*}
\caption{Construction and audit statistics for \dataset. PII types are sample-level semantic labels; evidence boxes are location-level visual items.}
\label{tab:dataset_audit}
\end{table}

Table~\ref{tab:dataset_audit} summarizes construction-audit coverage. All 832 samples include complete task, context, decision, masking, and PII-type metadata. Sample-level PII annotations cover medical, address, financial, identity, contact, credential, compensation, supplier/customer, and security-related signals; the largest localizable evidence categories are medical, financial, contact, address/location, and account or identity credentials.

\section{The \method{} Framework}

\method{} is an edge-side privacy and personalization layer between the live GUI and downstream agents. It arbitrates each outgoing observation inside a trusted user- or organization-controlled environment and refines auditable scenario-level privacy skills from validated feedback.

\subsection{Framework Overview}

\method{} has two coupled loops. In the online loop, the edge-side controller keeps raw screenshots local, extracts evidence, selects an exposure action through policy arbitration, and controls what leaves the device through SafeScreenshot construction. The evolution loop converts confirmations, cancellations, edits, and rejected actions into candidate skill updates that enter local memory after validation. Execution semantics remain fixed, and inspectable policy memory supports personalization.

\begin{figure*}[t]
\centering
\includegraphics[width=\textwidth,trim=58 0 54 30,clip]{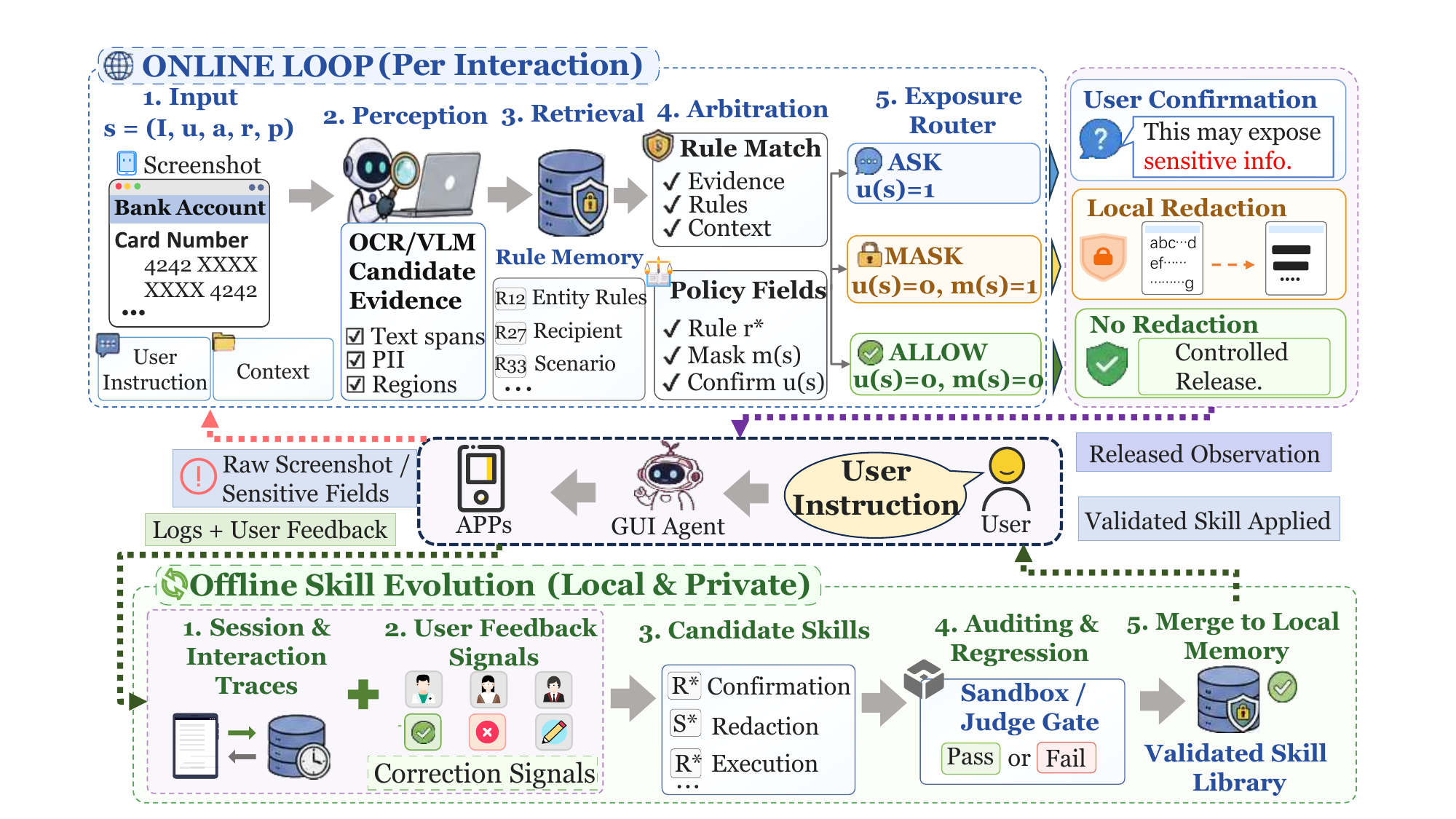}
\caption{\method{} architecture. The edge-side layer keeps raw screenshots local, extracts visual evidence, retrieves persona- and task-conditioned policy rules, and arbitrates an Allow/Mask/Ask exposure action. After arbitration, downstream agents receive allowed content, SafeScreenshot outputs, or confirmation requests. Validated user feedback updates auditable privacy skills.}

\label{fig:architecture}
\end{figure*}

Figure~\ref{fig:architecture} summarizes the pipeline. We next describe its four components.

\subsection{Local Perception and Evidence Extraction}

Perception extracts screen evidence for privacy reasoning before any raw screenshot leaves the device. It produces structured evidence for arbitration and SafeScreenshot construction.

For the current interaction, denoted by $s$, \method{} extracts an evidence set:
\begin{equation}
E(s)=\{e_1,e_2,\ldots,e_n\}.
\end{equation}
Each item may contain OCR text, field type, semantic description, confidence, and, when available, a bounding box. Localized evidence supports redaction. Non-localized evidence informs arbitration, following GUI parsing and grounding work \citep{yang2023,zheng2024}. Perception proposes evidence, and arbitration decides exposure.

\subsection{Rule-Grounded Policy Arbitration}

Arbitration is the fixed policy interface between local perception and the exposure action. \method{} grounds this interface in local policy memory:
\begin{equation}
C_t=\{r_1,r_2,\ldots,r_m\},
\end{equation}
where each entry is represented at the interface level as
\begin{equation}
r_j=(s_j,t_j,a_j,p_j,z_j).
\end{equation}
Here $s_j$ is the rule scope, covering application, persona or role, recipient boundary, and intended action; $t_j$ is the trigger evidence, such as OCR text, field type, semantic description, or a localizable region; $a_j$ is the recommended exposure action; $p_j$ is a conflict-resolution priority; and $z_j$ is a short rationale. This schema keeps policy conditions inspectable at the interface level.

For each interaction, \method{} builds a retrieval query from application, persona, recipient boundary, intended action, and local visual evidence $E(s)$. It retrieves a small candidate set $R_k(s)$ from $C_t$ using retrieval-augmented generation with local, user-specific memory \citep{lewis2020,asai2024}. Candidate rules are ranked by facet match and evidence compatibility, with recipient, action, and persona matches weighted above generic field-type matches.

The controlled arbitrator selects an applicable entry $r^\ast$ from $R_k(s)$ and emits a masking variable $m(s)$ and a confirmation variable $u(s)$ as locked fields. More specific persona-recipient-action rules override generic rules. When the retrieved set lacks a sufficiently applicable rule, the arbitrator falls back to conservative boundary tests over task necessity, recipient trust, and sensitive evidence. The final policy mapping is deterministic:
\begin{equation}
\mathrm{Decision}(s)=
\begin{cases}
\text{Ask}, & u(s)=1,\\
\text{Mask}, & u(s)=0 \wedge m(s)=1,\\
\text{Allow}, & \text{otherwise}.
\end{cases}
\end{equation}

Confirmation has priority over masking, and masking has priority over allowing. The Allow/Mask/Ask interface remains fixed and auditable as policy memory evolves.

\subsection{SafeScreenshot Construction}

SafeScreenshot construction turns the arbitration result into an observation boundary. Let $B(s)$ be the localizable regions from perception. Under Mask, \method{} uses the selected policy rule and local evidence to choose protected regions:
\begin{equation}
B^\ast(s)\subseteq B(s).
\end{equation}
It then constructs a mediated screenshot:
\begin{equation}
I_{\mathrm{safe}}=\mathrm{redact}(I,B^\ast(s)).
\end{equation}
The redaction operation can blur, mask, replace, or otherwise transform protected regions locally. Policy gating lets the same field remain visible, be masked, or trigger confirmation depending on the task, recipient, and user context. The mediated screenshot preserves task-relevant layout and non-sensitive content and weakens protected content.

Under Ask, \method{} pauses execution and returns a confirmation request with its rationale. Under Allow, the task proceeds through the controlled observation route with raw-screen access mediated by arbitration.

\subsection{Behavior-Driven Skill Evolution}

Skill evolution converts post-decision feedback into persistent local memory \citep{packer2023}. \method{} records traces:
\begin{equation}
\tau=[(s_1,d_1,y_1),\ldots,(s_T,d_T,y_T)],
\end{equation}
where $s_t$ is the GUI state, $d_t$ the system decision, and $y_t$ contains downstream actions and feedback such as confirmations, cancellations, edits, rejected actions, retries, or explicit skill instructions.

From these traces, \method{} extracts correction signals $G=\{g_1,\ldots,g_l\}$ encoding reusable constraints, such as masking fields for external recipients or confirming irreversible actions. Candidate updates $\Delta C_t$ may revise skill conditions, insert steps, prohibit actions, or require confirmation. A fixed scorer guides hill-climbing over textual skills, and the selected candidate enters a serial audit:
\begin{equation}
C_{t+1}=\mathrm{AuditMerge}(C_t,\Delta C_t).
\end{equation}

The audit has three gates. First, a schema and coverage gate rejects malformed skills, missing core conditions, and keyword-only rewrites that lack binding to an app, action, recipient, or sensitive evidence. Second, a fixed text scorer evaluates rule coverage, held-out decisions, unsafe allows, and concise skill quality. Third, an LLM-Judge / sandbox gate audits confirmation timing, safety rationale, and executable state flow for downstream agents. This final gate catches skills that mention the right privacy rule and still produce unsafe action order or incomplete flow.

In our implementation, evolution is controlled hill climbing over text memory: mutations are accepted only when they improve the current best fixed-scorer skill and pass the serial audit. The process follows LLM-based textual optimization and recent skill-evolution systems \citep{yang2024,jiang2026}.

\section{Experimental Setup}

\subsection{Research Questions and Evaluation Map}

We organize evaluation around three experiment-level questions. E1 measures whether the local privacy pipeline can perceive sensitive evidence, improve retrieval-grounded arbitration, and construct SafeScreenshots. E2 diagnoses whether simpler deployment routes can replace local policy memory, and E3 evaluates whether user feedback can improve scenario-level skills. L1--L3 denote pipeline-layer diagnostics: L2 evaluates full-benchmark policy arbitration, while L1 and L3 audit visual triage, extraction, and SafeScreenshot construction. Server-side replay and L1/L3 details are reported in the appendices.

\begin{table}[t]
\centering
\small
\setlength{\tabcolsep}{3pt}
\renewcommand{\arraystretch}{1.08}
\begin{tabularx}{\linewidth}{@{}lX@{}}
\toprule
\textbf{ID} & \textbf{Focus and evidence}\\
\midrule
E1 & Local pipeline: L1 OCR coverage, L2 Open vs.\ Closed arbitration on 832 rows, and L3 SafeScreenshot audit.\\
E2 & Policy memory: static, cloud, and edge-cloud routes under one evaluator.\\
E3 & Skill evolution: five controlled scenarios with fixed scorer and sandbox gate.\\
\bottomrule
\end{tabularx}
\caption{Experiment map. E1 combines component diagnostics with the benchmark L2 policy protocol. E2 uses the benchmark policy protocol. E3 evaluates controlled skill evolution.}
\label{tab:eval_map}
\end{table}

\begin{table*}[t]
\centering
\small
\begin{tabular*}{\textwidth}{@{}>{\centering\arraybackslash}p{.16\textwidth}@{\extracolsep{\fill}}>{\centering\arraybackslash}p{.25\textwidth}cccc@{}}
\toprule
\textbf{Category} & \textbf{System} & \textbf{Raw screen} & \textbf{Context} & \textbf{Local policy} & \textbf{Skill update}\\
\midrule
\multirow{1}{*}{Detector} & Static Regex / Presidio & No & none & No & No\\
\cmidrule(l){1-6}
\multirow{1}{*}{Static memory} & Static Policy Corpus & No & task/persona & static & No\\
\cmidrule(l){1-6}
\multirow{3}{*}{Cloud VLM} & Cloud VLM Minimal & Yes & minimal & No & No\\
& Cloud VLM Persona & Yes & persona & No & No\\
& Cloud Context-Complete Persona & Yes & full non-label & No & No\\
\cmidrule(l){1-6}
\multirow{1}{*}{Routing} & EdgeClaw-ClawXRouter & mixed & route features & No & No\\
\cmidrule(l){1-6}
\multirow{1}{*}{Local ablation} & \method{} Local VLM & No & full non-label & No & No\\
\cmidrule(l){1-6}
\multirow{2}{*}{Ours} & Policy-grounded \method{} & No & full non-label & static & No\\
& \method{} with Skills & No & full non-label & rules+skills & controlled\\
\bottomrule
\end{tabular*}
\caption{Compared systems grouped by baseline class. Full non-label context excludes expected decisions, PII/risk tags, and policy-memory answers.}
\label{tab:systems}
\end{table*}

\subsection{Compared Systems}

We group baselines into detector, static-memory, cloud-reasoning, and routing classes \citep{siyan2025}. Table~\ref{tab:systems} summarizes input context, raw-screen exposure, local policy access, and skill updates. E2 is a benchmark-wide diagnostic over 832 canonical rows, with systems varying exposure path, context, and access to policy memory. Cloud routes may see raw screenshots; expected labels, PII/risk tags, and policy-memory answers are withheld. Policy-grounded \method{} isolates static local policy memory, and skill-augmented \method{} is evaluated in E3.

\subsection{Static Decision Protocol and Metrics}
\label{sec:metrics}

E1 and E2 use a static single-step protocol. Inputs include the screenshot, instruction, candidate intent, app/recipient context, and persona; expected labels, risk tags, and PII categories are evaluator-only. Systems output policy and route fields, and the evaluator recomputes the final Allow/Mask/Ask decision.

The decision governs exposure of the current visual observation. Downstream task success and action authorization are outside the scoring target. The evaluator applies a conservative priority: Ask overrides Mask, and Mask overrides Allow. Ask covers consent-sensitive or irreversible exposure, Mask covers redactable sensitive content, and Allow requires task necessity and persona-policy compatibility.

We report joint decision accuracy, Mask F1, Ask recall, leak rate, over-protection, and raw-upload rate. For error analysis, we define \emph{Leak} ..., \emph{OverProtect} ..., and \emph{AskMiss} ... . Region-level masking errors, such as masking the wrong protected region, are treated as SafeScreenshot diagnostics and reported in Appendix~\ref{app:component_checks}.

\subsection{Layer-wise Diagnostics}

E1 also includes layer-wise diagnostics. L2 uses all 832 policy rows; L1/L3 audit visual PII triage, sensitive-item extraction, and strict OCR-based SafeScreenshot construction on applicable subsets. Appendix~\ref{app:component_checks} gives denominators, QA flags, and artifacts.

\subsection{Self-Evolution Evaluation}

E3 evaluates adaptation across five scenario families: iCloud cleanup, app permission, high-value transfer, schedule merge, and driving mode. Each scenario contains a task goal, three initial skills, user corrections, target constraints, and held-out decision and behavior tests. Corrections are evolution inputs; constraints and tests are fixed for scoring.

We use five scenarios, three start conditions, three random seeds, and 20 evolution iterations. Each skill is scored as:
\begin{equation}
\begin{aligned}
J(\theta) ={}& \lambda_1\,\mathrm{RuleCoverage}
+ \lambda_2\,\mathrm{TestDecisionAcc}\\
&+ \lambda_3\,\mathrm{SafetyPenalty}
+ \lambda_4\,\mathrm{SkillQuality},
\end{aligned}
\label{eq:skill_objective}
\end{equation}
where $\lambda_1=0.40$, $\lambda_2=0.30$, $\lambda_3=0.20$, and $\lambda_4=0.10$. Weights are fixed before evaluation. Rule coverage and held-out correctness dominate, safety is enforced by filters, and SkillQuality favors concise, inspectable skills. Rule-based caps or rejection handle malformed skills, critical safety violations, missing core rules, and unsafe allow decisions. The LLM-Judge audits privacy conditions, confirmation timing, and executable state transitions for downstream agents, following recent GUI evaluation practice \citep{zheng2023,lin2025}. We use BestScore@20 as the primary evolution metric and report behavior accuracy, unsafe action rate, and correction compliance as skill-use checks.

\section{Results and Analysis}
\subsection{Local Arbitration and SafeScreenshot Checks}

\begin{table}[t]
\centering
\small
\setlength{\tabcolsep}{4pt}
\begin{tabular}{@{}lrrrr@{}}
\toprule
\textbf{Check} & \textbf{Score} & \textbf{Mask R.} & \textbf{Allow R.} & \textbf{Ask R.}\\
\midrule
L1 OCR coverage & 967 & \multicolumn{1}{c}{--} & \multicolumn{1}{c}{--} & \multicolumn{1}{c}{--}\\
L2 Open & $533{\pm}10$ & $770{\pm}7$ & $308{\pm}12$ & $113{\pm}22$\\
L2 Closed & $717{\pm}7$ & $804{\pm}6$ & $683{\pm}18$ & $377{\pm}32$\\
L3 strict redaction & 733 & \multicolumn{1}{c}{--} & \multicolumn{1}{c}{--} & \multicolumn{1}{c}{--}\\
\bottomrule
\end{tabular}
\caption{E1 layer diagnostics. Values are reported in $10^{-3}$: Score is normalized exact OCR text coverage for L1, policy accuracy for L2, and strict no-flag rate for L3. Only L2 Open and L2 Closed are directly comparable policy-decision results; component details are in Appendix~\ref{app:component_checks}.}
\label{tab:l2}
\end{table}

Table~\ref{tab:l2} shows the E1 evidence chain. The core L2 result improves policy accuracy from $0.533{\pm}0.010$ to $0.717{\pm}0.007$. L1 reaches $0.967$ normalized exact OCR text coverage, and L3 passes the strict no-flag audit on most visible-PII cases. The L2 gain is contextual: Allow and Ask recall rise sharply while Mask recall remains high. Earlier frozen-interface diagnostics are reported in Appendix~\ref{app:c3c5}.

\subsection{Baseline Comparison}

\begin{table*}[!t]
\centering
\small
\setlength{\tabcolsep}{5pt}
\begin{tabular*}{\textwidth}{@{}l@{\extracolsep{\fill}}lccccc@{}}
\toprule
\textbf{Method} & \textbf{Family} & \textbf{Acc.} & \textbf{Mask F1} & \textbf{Leak} & \textbf{Ask R.} & \textbf{Raw up.}\\
\midrule
Policy-grounded \method{} & policy memory & $717{\pm}7$ & $819{\pm}14$ & $196{\pm}6$ & $377{\pm}32$ & 0\\
Static Regex & pattern detector & $557{\pm}0$ & $645{\pm}0$ & $283{\pm}0$ & $0{\pm}0$ & 0\\
Static Policy Corpus & static memory & $254{\pm}0$ & $413{\pm}0$ & $637{\pm}0$ & $438{\pm}0$ & 0\\
Cloud Minimal & cloud reasoning & $308{\pm}6$ & $359{\pm}14$ & $709{\pm}13$ & $613{\pm}45$ & 1000\\
Cloud Persona & cloud reasoning & $370{\pm}4$ & $472{\pm}7$ & $540{\pm}9$ & $263{\pm}13$ & 1000\\
Cloud Full-Context & cloud reasoning & $166{\pm}5$ & $83{\pm}6$ & $956{\pm}4$ & $971{\pm}19$ & 1000\\
EdgeClaw-ClawXRouter$^\dagger$ & exposure routing & $401{\pm}6$ & $480{\pm}3$ & $552{\pm}1$ & $529{\pm}19$ & 159\\
\bottomrule
\end{tabular*}
\caption{Architectural diagnostic of privacy routes. All values are reported in $10^{-3}$. Values with $\pm$ summarize repeated runs; routing exposure is deterministic or route-derived. Policy-grounded \method{} includes local policy memory but excludes skill evolution; cloud routes are rule-free reasoning controls. Cloud Full-Context denotes Cloud Context-Complete Persona. Leak is expected Mask predicted as non-Mask. $^\dagger$EdgeClaw-ClawXRouter is evaluated as a route-level comparison; details are in Appendix~\ref{app:baseline_details}.}
\label{tab:main}
\end{table*}

Static Regex is the strongest non-\method{} baseline by accuracy, reaching $0.557$ despite having no task, persona, recipient, or policy memory. Visible-pattern detection provides a competitive lower bound and is close to the ungrounded local VLM ablation in Table~\ref{tab:l2}. Its limited contextual policy semantics produce zero Ask recall, conflate task-necessary exposure with over-protection, and leave a leak rate of $0.283$. Policy-grounded \method{} improves over this strongest baseline by $0.160$ absolute accuracy, from $0.557$ to $0.717$. It also increases Mask F1 by $0.174$ and reduces leak by $0.087$.

Cloud and routing controls show different route-level behaviors. Cloud routes upload every raw screenshot, and Cloud Full-Context is dominated by Ask predictions, with high Ask recall but only $0.166{\pm}0.005$ accuracy. EdgeClaw-ClawXRouter reduces raw-cloud routing to 15.9\%. As an exposure-routing proxy, it reaches $0.401{\pm}0.006$ accuracy. Table~\ref{tab:main} shows that pattern detectors catch many visible spans, cloud routes tend to over-confirm or expose raw screens, and routing proxies leave final contextual arbitration unresolved. Appendix~\ref{app:baseline_details} reports rerun checks.

\subsection{Skill Evolution and Behavior Improvement}
\begin{figure}[tb]
\centering
\includegraphics[width=\linewidth]{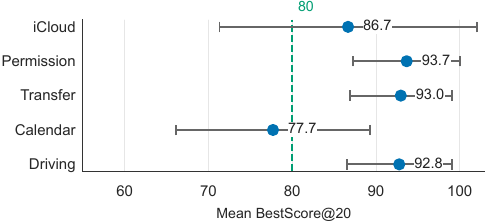}
\caption{Skill-use checks in five controlled scenarios. The plot reports downstream behavior after text-level skill evolution.}
\label{fig:e4_behavior}
\end{figure}

Experiment 3 focuses on controlled skill evolution. The first question is whether correction-driven search can produce high-quality textual skills under a fixed scorer. The second is whether a separate sandbox gate rejects candidates whose text looks good and whose execution remains unsafe. Figure~\ref{fig:e4_behavior} checks whether selected skills also improve downstream behavior.

Across five controlled scenarios, BestScore@20 averages 88.76, with scores ranging from 77.70 to 93.67. The primary evolution result is that user corrections can be distilled into textual skills that satisfy fixed coverage, held-out decision, safety-penalty, and quality constraints in designed settings.

Figure~\ref{fig:e4_behavior} shows that evolved skills improve held-out behavior decisions and reduce unsafe actions. Compliance remains imperfect (76.19\% for transfer and 60.87\% for driving), so a high-scoring textual skill can still miss evidence, wording, or order constraints when used by a downstream agent. This gap motivates the post-optimization sandbox filter below.

\subsection{Sandbox Validation}

Sandbox validation serves as the post-evolution safety audit. It checks confirmation timing, rationale, and executable state flow before a candidate can be treated as reusable memory.

\begin{figure}[tb]
\centering
\includegraphics[width=\linewidth]{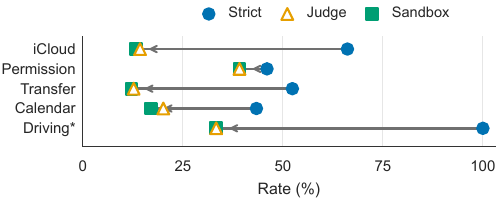}
\caption{Post-optimization filtering from strict text scoring to the LLM-Judge / sandbox gate. Rejections mainly involve state flow, safety reasoning, and rule alignment.}
\label{fig:sandbox_judge}
\end{figure}

Figure~\ref{fig:sandbox_judge} reports post-optimization filtering from strict text scoring to the flow-level sandbox gate. The drop is expected because the two gates test different properties: iCloud moves from 66.14\% strict pass to 13.23\% sandbox pass, and transfer moves from 52.38\% to 12.17\%. Candidate skills can satisfy textual privacy constraints and still fail executable flow-level safety.

Thus, low sandbox pass rates show why deployment-oriented filtering is necessary after text-level skill search. The LLM-Judge provides an audit signal for reusable memory selection; full counts and diagnostics are in the appendix.

\subsection{Error Analysis}

The remaining failures are primarily contextual boundary errors. In the 832-example joint-field diagnostic, the finalized arbitration interface reaches 0.7127 joint accuracy, with 593 correct cases and 239 joint errors. The three largest categories are protected content allowed, unnecessary masking, and confirmation replaced by masking, accounting for 79.5\% of joint errors.

These errors cluster around persona, application, action, data type, and recipient boundaries: whether exposure is task-necessary, which recipient is trusted, and when confirmation is required. Retrieval diagnostics often find a related rule for the wrong boundary, suggesting boundary confusion and rule-applicability failures.

\section{Conclusion}
This paper frames GUI-agent screenshot privacy as exposure arbitration and presents \method{}, an edge-side framework that decides what should be allowed, masked, or confirmed under task, recipient, and persona context. This keeps privacy handling explicit while preserving useful visual context for downstream agents through allowed, masked, or confirmed observations at runtime. Across \dataset{}, policy-memory-free or route-only baselines struggle with these boundaries and often expose raw screenshots. The skill-evolution study shows a narrower but useful result: in designed scenarios, user corrections can become reusable privacy skills, and sandbox checks catch candidates that still fail flow-level safety.

\section*{Limitations}

\dataset{} is a sanitized benchmark for GUI privacy decisions, built from reconstructed scenarios. It covers multiple personas, tasks, recipients, and UI perturbations, with personas serving as author-designed abstractions for workflow and policy variation. Real-user and organization-level validation remain future work. Our skill-evolution study is scenario-level: it shows that corrections can be distilled into reusable privacy skills in designed settings, while longer-term personalization requires broader user studies. The deployment scope is trusted edge-side execution in a user or organization environment such as a workstation or private server. These limits bound the empirical claims while preserving the core contribution: \method{} reduces raw screenshot exposure before downstream agents receive GUI observations.

\section*{Ethics Statement}

\dataset{} is built from simulated and sanitized GUI scenarios, with real user logs, private chats, medical records, payment credentials, identity documents, and personal identifiers outside the release scope. Released artifacts should be reviewed for accounts, contacts, organization identifiers, QR codes, credentials, and reverse-identifiable information; release-eligible assets should pass this review, while other assets should be withheld or replaced by metadata-only records. \method{} reduces raw screenshot exposure through local evidence extraction, policy arbitration, and visual redaction; cloud VLM and edge-cloud routes are diagnostic baselines. The intended use is GUI-agent privacy research, with high-risk medical, financial, identity, enterprise-compliance, and irreversible actions requiring separate professional or legal review. Evolved skills may encode user preferences and should remain inspectable; LLM judges, masking, and sandbox validation provide risk-reduction checks.
\bibliography{main}

\clearpage
\appendix
\input{appendix_draft}

\end{document}

%% file: appendix_draft.tex
% Appendix draft for acl_latex.tex.
% Suggested use: paste after \appendix, or add \input{appendix_draft} after \appendix.
% This file contains appendix body only; the main paper supplies \appendix and \bibliography.

\section{Evaluation Protocol and Dataset Details}
\label{app:protocol}

Prediction methods receive only fields that would be available to a deployed privacy mediator: the screenshot, application context, persona context, user instruction, and candidate agent intent. Audit-only fields, including PII category, risk tags, expected labels, and author audit notes, are used only for offline scoring and diagnostics. This separation prevents the evaluator from leaking expected policy information into the decision process.

P-GUI-Evo contains 832 controlled GUI privacy samples built from reconstructed UI layouts, LLM-assisted scenario drafting, typed sanitized sensitive-field injection, and audit filtering. Tables and prose use the display labels Allow (314), Mask (438), and Ask (80). The benchmark includes three personas and three buckets, summarized in Table~\ref{tab:app_dataset_distribution}. D1 contains clean in-distribution cases, D2 contains UI and task variants, and D3 contains stress cases with boundary conditions and UI perturbations. The final set also keeps controlled variants around shared scenario cores: structural UI shifts (\texttt{ss}), visual shifts (\texttt{vs}), and lexical or semantic task shifts (\texttt{ls}, lowercase L-s).

\begin{table}[t]
\centering
\small
\begin{tabular*}{\linewidth}{@{}p{.54\linewidth}@{\extracolsep{\fill}}rr@{}}
\hline
\textbf{Partition} & \textbf{Count} & \textbf{Share}\\
\hline
\multicolumn{3}{l}{\textit{Policy label}}\\
Mask & 438 & 52.6\%\\
Allow & 314 & 37.7\%\\
Ask & 80 & 9.6\%\\
\hline
\multicolumn{3}{l}{\textit{Persona}}\\
User A & 234 & 28.1\%\\
User B & 246 & 29.6\%\\
User C & 352 & 42.3\%\\
\hline
\multicolumn{3}{l}{\textit{Bucket}}\\
D1 basic & 174 & 20.9\%\\
D2 generalization & 546 & 65.6\%\\
D3 stress & 112 & 13.5\%\\
\hline
\multicolumn{3}{l}{\textit{Scenario variant}}\\
Base screenshot & 250 & 30.0\%\\
Structural/UI shift (\texttt{ss}) & 201 & 24.2\%\\
Visual shift (\texttt{vs}) & 208 & 25.0\%\\
Lexical/semantic shift (\texttt{ls}, lowercase L) & 173 & 20.8\%\\
\hline
\end{tabular*}
\caption{Dataset distribution used by the static policy-decision experiments. Percentages are computed over 832 samples. Scenario variants preserve the same underlying privacy boundary while changing the UI structure, visual rendering, or task wording.}
\label{tab:app_dataset_distribution}
\end{table}

\section{L2 Retrieval-Augmented Arbitration Diagnostics}
\label{app:c3c5}

The L2 Closed result in the main text reports the finalized retrieval-augmented policy arbitration interface on the canonical 832-sample benchmark. We include an earlier interface only to show how the final design changes the error profile; the main claims use the finalized interface.

Table~\ref{tab:app_c3_generalization} shows that relevant rules are usually present: the full-set top-k oracle reaches 0.9940. The remaining gap is mainly an arbitration gap, where related rules are retrieved but applied to the wrong policy boundary.

\begin{table}[h]
\centering
\footnotesize
\setlength{\tabcolsep}{3pt}
\begin{tabular*}{\linewidth}{@{}p{.37\linewidth}@{\extracolsep{\fill}}rrrr@{}}
\hline
\textbf{Split} & \textbf{N} & \textbf{Policy} & \textbf{Mask} & \textbf{Ask-R}\\
\hline
Preliminary pilot & 120 & 758 & 900 & 700\\
Preliminary held-out & 732 & 616 & 824 & 478\\
Preliminary full set & 832 & 609 & 821 & 400\\
\hline
\end{tabular*}
\caption{Preliminary-interface diagnostics across pilot, held-out, and full splits. Metric values are reported in $10^{-3}$. These diagnostics are included only to contextualize the finalized interface.}
\label{tab:app_c3_generalization}
\end{table}

\begin{table}[h]
\centering
\scriptsize
\setlength{\tabcolsep}{2.5pt}
\begin{tabular*}{\linewidth}{@{}l@{\extracolsep{\fill}}cccc@{}}
\hline
\textbf{Interface} & \textbf{Policy} & \textbf{Mask R.} & \textbf{Mask Acc.} & \textbf{Ask R.}\\
\hline
Preliminary interface & 609 & 821 & 722 & 400\\
Finalized interface & $717{\pm}7$ & $784{\pm}15$ & $904{\pm}3$ & $377{\pm}32$\\
\hline
\end{tabular*}
\caption{Preliminary versus finalized retrieval-augmented policy arbitration on the full 832-sample set. Values are reported in $10^{-3}$. The finalized interface is the setting used for the main result.}
\label{tab:app_c3_c5}
\end{table}

\paragraph{Mediator responsibility.}
The L2 Closed setting uses a deterministic mediator to choose among top-k rules. The closed pipeline is:
\begin{quote}
\small
query/context $\rightarrow$ local rule retrieval $\rightarrow$ deterministic mediator $\rightarrow$ local VLM visual evidence check.
\end{quote}
The deterministic mediator sets the selected rule id, mask flag, and confirmation flag. The local VLM checks whether the screenshot contains supporting visual evidence or an obvious contradiction. The mediator owns the locked policy fields, while the VLM reports visual evidence and contradiction flags. This responsibility split attributes the finalized-interface gain to rule-grounded arbitration with retrieval support and visual verification.

\section{Full Baseline Metrics and Route-Level Comparison}
\label{app:baseline_details}

The main text reports a route-oriented architectural diagnostic; Table~\ref{tab:app_full_baselines} gives the full score table. All systems are rescored against the canonical 832-row expected-decision table. Policy-grounded \method{} isolates rule-grounded static policy mediation, while cloud routes are rule-free controls. Skill recovery and skill-routing experiments are excluded from this diagnostic.

\begin{table*}[t]
\centering
\scriptsize
\setlength{\tabcolsep}{2pt}
\begin{tabular*}{\textwidth}{@{}p{.22\textwidth}@{\extracolsep{\fill}}rrrrrrrp{.06\textwidth}rrr@{}}
\hline
\textbf{Method} & \textbf{Acc.} & \textbf{M-P} & \textbf{M-R} & \textbf{M-F1} & \textbf{Leak} & \textbf{FPR} & \textbf{Ask-R} & \textbf{Route} & \textbf{Raw} & \textbf{Masked} & \textbf{Local}\\
\hline
Policy-grounded \method{} & 717 & 836 & 804 & 819 & 196 & 244 & 377 & local & 0 & 0 & 1000\\
Static Regex & 557 & 587 & 717 & 645 & 283 & 525 & 0 & local & 0 & 0 & 1000\\
Static Policy Corpus & 254 & 480 & 363 & 413 & 637 & 405 & 438 & local & 0 & 0 & 1000\\
Cloud Minimal & 308 & 471 & 294 & 359 & 709 & 385 & 613 & cloud & 1000 & 0 & 0\\
Cloud Persona & 370 & 490 & 470 & 472 & 540 & 535 & 263 & cloud & 1000 & 0 & 0\\
Cloud Context-Complete Persona & 166 & 724 & 48 & 83 & 956 & 22 & 971 & cloud & 1000 & 0 & 0\\
EdgeClaw-ClawXRouter & 401 & 512 & 448 & 480 & 552 & 484 & 529 & hybrid & 159 & 460 & 381\\
\hline
\end{tabular*}
\caption{Full route-diagnostic metrics on the 832-sample benchmark with explicit routing exposure. Values are reported in $10^{-3}$. Main policy metrics follow the rerun summary where available; routing columns report exposure-mode rates on the same scale.}
\label{tab:app_full_baselines}
\end{table*}

\paragraph{Repeated-run stability.}
Repeated runs on the same 832 canonical samples show the same pattern: policy-grounded \method{} remains the strongest local policy route, cloud routes remain conservative and Ask-heavy, and EdgeClaw-ClawXRouter is best interpreted as a route-level comparison rather than a contextual arbitration system.

\begin{table}[h]
\centering
\scriptsize
\setlength{\tabcolsep}{2pt}
\resizebox{\linewidth}{!}{%
\begin{tabular}{@{}lrrrr@{}}
\hline
\textbf{Method} & \textbf{Acc.} & \textbf{Mask F1} & \textbf{Leak} & \textbf{Ask-R}\\
\hline
Regex & $557{\pm}0$ & $645{\pm}0$ & $283{\pm}0$ & $0{\pm}0$\\
Policy-grounded & $717{\pm}7$ & $819{\pm}14$ & $196{\pm}6$ & $377{\pm}32$\\
Static Policy & $254{\pm}0$ & $413{\pm}0$ & $637{\pm}0$ & $438{\pm}0$\\
Cloud Min. & $308{\pm}6$ & $359{\pm}14$ & $709{\pm}13$ & $613{\pm}45$\\
Cloud Pers. & $370{\pm}4$ & $472{\pm}7$ & $540{\pm}9$ & $263{\pm}13$\\
Cloud Full & $166{\pm}5$ & $83{\pm}6$ & $956{\pm}4$ & $971{\pm}19$\\
EdgeClaw & $401{\pm}6$ & $480{\pm}3$ & $552{\pm}1$ & $529{\pm}19$\\
\hline
\end{tabular}
}
\caption{Three-run stability checks for valid repeated controls. Values are reported in $10^{-3}$ as mean $\pm$ sample standard deviation on the same 832 canonical samples.}
\label{tab:app_baseline_stability}
\end{table}

Cloud Context-Complete Persona is a rule-free cloud control. It sees the raw screenshot plus app, command, agent intent, recipient context, and persona profile; expected labels, PII/risk tags, policy-memory answers, and label-revealing identifiers are withheld. Its high leak value mostly comes from expected Mask samples predicted as Ask, with few direct Allow predictions, indicating over-confirmation as the dominant behavior.

\begin{table}[h]
\centering
\footnotesize
\setlength{\tabcolsep}{4pt}
\begin{tabular*}{\linewidth}{@{}l@{\extracolsep{\fill}}rrr@{}}
\hline
\textbf{Expected} & \textbf{Pred Allow} & \textbf{Pred Mask} & \textbf{Pred Ask}\\
\hline
Allow & 44 & 7 & 263\\
Mask & 2 & 21 & 415\\
Ask & 3 & 1 & 76\\
\hline
\end{tabular*}
\caption{Cloud Context-Complete Persona confusion summary. The baseline predicts Ask for 754/832 samples, indicating over-confirmation as its dominant arbitration behavior.}
\label{tab:app_cloud_context_confusion}
\end{table}

Table~\ref{tab:app_cloud_context_confusion} explains the low Cloud Full-Context score. The model rarely directly allows protected cases, but it routes most policy decisions to confirmation. This conservative behavior is safer than unconstrained Allow, yet it fails the benchmark's policy-arbitration objective because many cases require a task-specific Mask or Allow decision.

\begin{table*}[t]
\centering
\scriptsize
\setlength{\tabcolsep}{3pt}
\begin{tabular*}{\textwidth}{@{}p{.22\textwidth}@{\extracolsep{\fill}}p{.13\textwidth}p{.10\textwidth}p{.28\textwidth}p{.20\textwidth}@{}}
\hline
\textbf{Cloud baseline} & \textbf{Image} & \textbf{Temp.} & \textbf{Context} & \textbf{Output handling}\\
\hline
Minimal & raw & 0.0 & app + command & parsed JSON policy fields\\
Persona & raw & 0.0 & app + command + persona profile & parsed JSON policy fields\\
Context-Complete Persona & raw & 0.0 & app; command; intent; recipient context; persona & parsed JSON policy fields\\
\hline
\end{tabular*}
\caption{Cloud baseline configuration. The table reports the information exposed to each rule-free cloud control; all expected labels and analysis-only metadata are withheld.}
\label{tab:app_cloud_config}
\end{table*}

EdgeClaw-ClawXRouter emits routing states for local, masked-cloud, and raw-cloud exposure. Its safety objective differs from \dataset{}: it decides whether an observation should remain local, be masked before cloud use, or be sent raw, whereas our benchmark asks for a contextual exposure policy. Table~\ref{tab:app_edgeclaw_mapping} maps its routes to the nearest Allow/Mask/Ask labels only for route-level comparison. In the original routing output, 46.0\% of samples are masked before cloud upload and 38.1\% remain local.

\begin{table*}[t]
\centering
\small
\setlength{\tabcolsep}{4pt}
\begin{tabular*}{\textwidth}{@{}l@{\extracolsep{\fill}}p{.25\textwidth}p{.22\textwidth}p{.16\textwidth}rr@{}}
\hline
\textbf{State} & \textbf{Route label} & \textbf{Upload mode} & \textbf{Mapped decision} & \textbf{Count} & \textbf{Share}\\
\hline
S1 & \texttt{cloud\_raw} & \texttt{raw\_image} & Allow & 132 & 15.9\%\\
S2 & \texttt{cloud\_after\_mask} & \texttt{masked\_image} & Mask & 383 & 46.0\%\\
S3 & \texttt{local\_only} & \texttt{local\_only} & Ask & 317 & 38.1\%\\
\hline
\end{tabular*}
\caption{Route-level comparison for EdgeClaw-ClawXRouter. Counts sum to the 832-sample benchmark. The mapped decision is used only to place the router in the shared policy-decision table; the upload mode preserves the hybrid edge-cloud behavior.}
\label{tab:app_edgeclaw_mapping}
\end{table*}

\section{L1/L3 Component Checks and SafeScreenshot Audit}
\label{app:component_checks}

The L1 and L3 checks are module-level diagnostics. L1 uses visual triage and sensitive-item extraction, and the main-table L1 score reports OCR text coverage by matching normalized gold sensitive-text values against OCR output. L3 evaluates automatic SafeScreenshot construction: sensitive items are extracted, OCR localizes matching text, and a strict remasking rule masks matched regions while producing conservative audit flags.

\begin{figure*}[t]
\centering
\includegraphics[width=\textwidth]{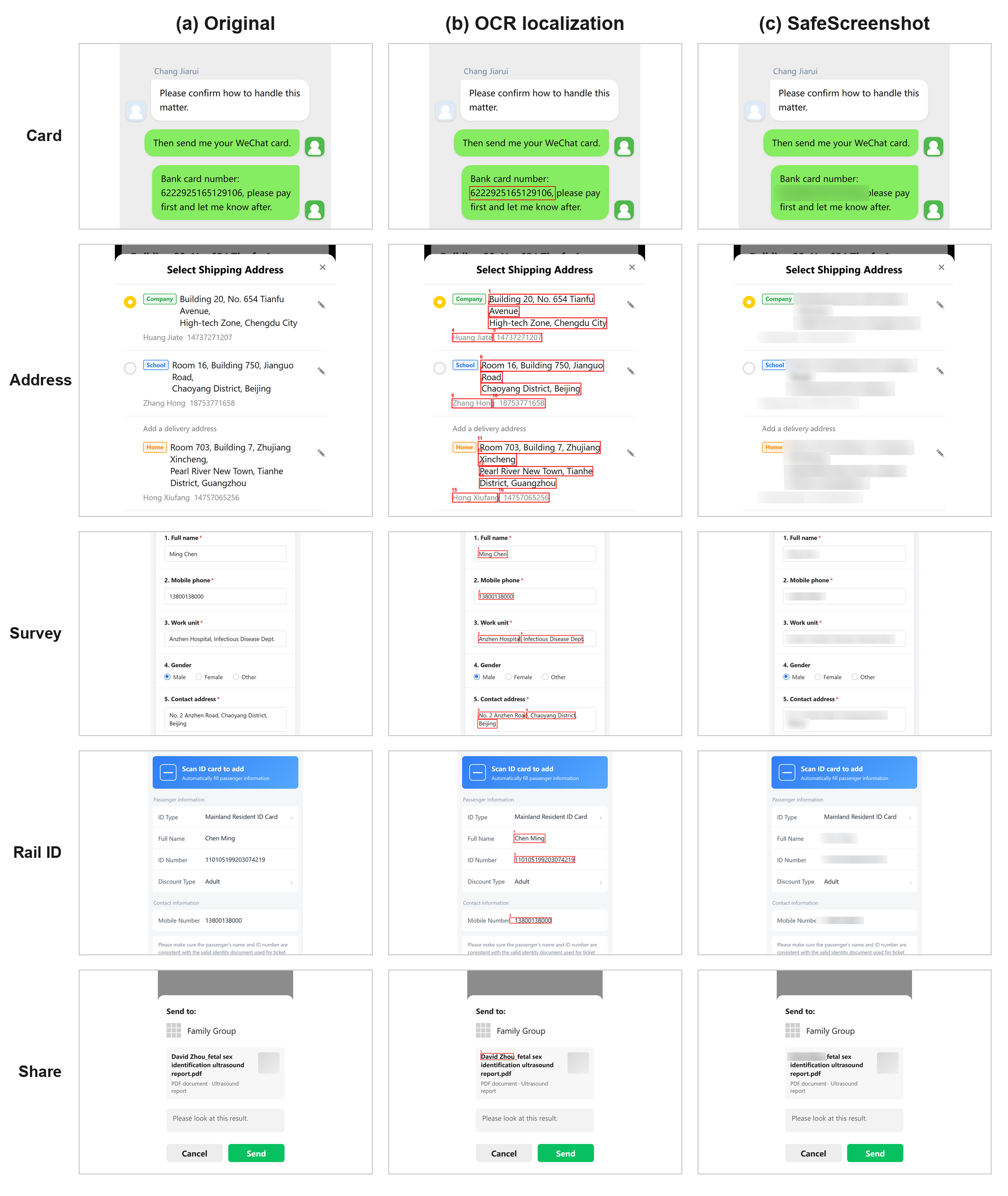}
\caption{Qualitative SafeScreenshot examples from the strict L3 audit. The first row shows a synthetic bank-card span; the second row shows an address-list UI with multiple address and phone fields. OCR localizes matched sensitive text, and SafeScreenshot masks those spans while preserving task-relevant structure such as address-category tags. Both examples are no-flag audit cases; all shown identifiers and addresses are synthetic.}
\label{fig:app_l3_safescreenshot_example}
\end{figure*}

\begin{table}[h]
\centering
\footnotesize
\setlength{\tabcolsep}{3pt}
\begin{tabular*}{\linewidth}{@{}p{.42\linewidth}@{\extracolsep{\fill}}rp{.36\linewidth}@{}}
\hline
\textbf{Check} & \textbf{N} & \textbf{Result}\\
\hline
Visual triage & 832 & 832 ok; 620 visible-PII\\
L1 OCR text coverage & 736 & 1,446/1,496 exact; 99.96\% char recall\\
PII-type audit & 251 & reviewed for PII type\\
Visible-PII L3 set & 558 & 466 mask + 92 allow\\
Sensitive-item extraction & 558 & 558 ok; zero empty rows\\
\hline
\end{tabular*}
\caption{L1/L3 diagnostic denominators. These audits support benchmark checking as component-level quality signals.}
\label{tab:app_l1_l3_scope}
\end{table}

Table~\ref{tab:app_l3_strict} reports the strict L3 run used in the main text. It matches OCR boxes to extracted sensitive-item text or token-like values; broad type-keyword expansion and global regex fallback are used only in the recall-first diagnostic variant. The no-flag count is an automatic audit status for clear leakage or overmasking checks. Manual inspection verified that the strict run removed the broad overmasking patterns observed in the recall-first run.

\begin{table}[h]
\centering
\footnotesize
\setlength{\tabcolsep}{3pt}
\begin{tabular*}{\linewidth}{@{}l@{\extracolsep{\fill}}r@{}}
\hline
\textbf{Strict SafeScreenshot audit} & \textbf{Count}\\
\hline
Rows & 558\\
Mask generated & 475\\
Automatic no-flag cases & 409\\
Flagged for targeted audit & 149\\
Missed sensitive-token flags & 145\\
No OCR match flags & 83\\
Overmask-suspect flags & 5\\
\hline
\end{tabular*}
\caption{Strict SafeScreenshot construction audit on 558 visible-PII screenshots. Flags are conservative automatic audit signals and may overlap.}
\label{tab:app_l3_strict}
\end{table}

\begin{table}[h]
\centering
\footnotesize
\setlength{\tabcolsep}{3pt}
\begin{tabular*}{\linewidth}{@{}l@{\extracolsep{\fill}}rr@{}}
\hline
\textbf{Variant} & \textbf{Mask generated} & \textbf{No-flag cases}\\
\hline
Recall-first & 542/558 & 338/558\\
Strict & 475/558 & 409/558\\
\hline
\end{tabular*}
\caption{Recall-first versus strict remasking. Recall-first masks more text but triggers more audit flags; strict is the reported setting.}
\label{tab:app_l3_variants}
\end{table}

\section{Skill Evolution Scenario Protocol and Full Results}
\label{app:e4_full}

Experiment 3 evaluates scenario-level adaptation. A scenario includes a task goal, initial skills, user corrections, target constraints, and held-out behavior tests. The update target is a validated skill for downstream behavior; the static policy memory used for arbitration remains fixed in this experiment.

\begin{table*}[t]
\centering
\footnotesize
\setlength{\tabcolsep}{4pt}
\begin{tabular*}{\textwidth}{@{}p{.14\textwidth}@{\extracolsep{\fill}}p{.36\textwidth}p{.11\textwidth}rp{.25\textwidth}@{}}
\hline
\textbf{Scenario} & \textbf{Task boundary} & \textbf{Optimization} & \textbf{Tests} & \textbf{Evaluation note}\\
\hline
iCloud cleanup & Avoid deleting unsynced, favorited, or irreversible local media. & 9 starts, 20 iters & 8 & Full controlled setting.\\
App permission & Replace automatic permission approval with minimal, contextual authorization. & 9 starts, 20 iters & 8 & Full controlled setting.\\
High-value transfer & Enforce amount thresholds, liveness, payee checks, and final confirmation. & 9 starts & 15 & Valid-run subset with fixed behavior tests.\\
Calendar merge & Merge calendars while respecting lunch blocks, buffers, attendees, time zones, and confirmation. & 9 starts, 20 iters & 8 & Full controlled setting.\\
Driving mode & Minimize driver attention and prevent sensitive or non-urgent message disclosure. & 9 starts & 8 & Diagnostic subset with fixed behavior tests.\\
\hline
\end{tabular*}
\caption{Skill-evolution protocol by scenario. Evaluation notes distinguish full controlled settings from valid diagnostic subsets.}
\label{tab:app_skill_protocol}
\end{table*}

\begin{table*}[t]
\centering
\scriptsize
\setlength{\tabcolsep}{4pt}
\begin{tabular*}{\textwidth}{@{}p{.21\textwidth}@{\extracolsep{\fill}}rrrrrrr@{}}
\hline
\textbf{Scenario} & \textbf{Best@20} & \textbf{Tests} & \textbf{Base Acc.} & \textbf{Evo Acc.} & \textbf{Base Unsafe} & \textbf{Evo Unsafe} & \textbf{Compliance}\\
\hline
iCloud photo cleanup & 86.67 & 8 & 37.50 & 100.00 & 62.50 & 12.50 & 100.00\\
App permission & 93.67 & 8 & 25.00 & 100.00 & 75.00 & 0.00 & 100.00\\
High-value transfer & 92.96 & 15 & 20.00 & 100.00 & 80.00 & 0.00 & 76.19\\
Calendar merge & 77.70 & 8 & 12.50 & 100.00 & 87.50 & 0.00 & 95.65\\
Driving mode & 92.78 & 8 & 25.00 & 100.00 & 75.00 & 0.00 & 60.87\\
\hline
Macro average & 88.76 & -- & 24.00 & 100.00 & 76.00 & 2.50 & --\\
\hline
\end{tabular*}
\caption{Full behavior results for the five controlled skill-evolution scenarios. Best@20 is the mean best strict score within 20 iterations where applicable. Accuracy, unsafe action rate, and correction compliance are distinct metrics.}
\label{tab:app_skill_evolution}
\end{table*}

These results support controlled behavior improvement in five designed scenarios. The claim scope is scenario-level adaptation under fixed tests and curated feedback signals. In particular, compliance below 100\% in the transfer and driving scenarios means a behavior can become correct at a high level while still missing required evidence, confirmation timing, or ordering constraints.

\section{Sandbox and LLM-Judge Validation Details}
\label{app:sandbox}

The strict scorer checks text-level skill quality and rule coverage. The LLM-Judge / sandbox gate adds privacy-condition, confirmation-timing, safety-reasoning, and executable-flow checks.

\begin{table}[h]
\centering
\footnotesize
\setlength{\tabcolsep}{3pt}
\begin{tabular*}{\linewidth}{@{}p{.23\linewidth}@{\extracolsep{\fill}}rrrrp{.18\linewidth}@{}}
\hline
\textbf{Scenario} & \textbf{N} & \textbf{Strict} & \textbf{Judge} & \textbf{Sandbox} & \textbf{Reject}\\
\hline
iCloud & 189 & 66.14 & 14.29 & 13.23 & 80.00\\
App permission & 189 & 46.03 & 39.15 & 39.15 & 14.94\\
Transfer & 189 & 52.38 & 12.70 & 12.17 & 76.77\\
Calendar & 189 & 43.39 & 20.11 & 16.93 & 60.98\\
Driving & 9 & 100.00 & 33.33 & 33.33 & 66.67\\
\hline
\end{tabular*}
\caption{Sandbox validation pass rates. Values are percentages except candidate count. Driving uses a smaller diagnostic candidate set.}
\label{tab:app_sandbox}
\end{table}

The gap between strict and sandbox pass rates means the sandbox gate rejects candidates that look acceptable as text but fail flow-level constraints. Common failure modes include missing state transitions, unsafe confirmation ordering, and weak alignment with the scenario's target constraints.

\begin{table*}[t]
\centering
\scriptsize
\setlength{\tabcolsep}{4pt}
\begin{tabular*}{\textwidth}{@{}p{.22\textwidth}@{\extracolsep{\fill}}rrrrrrr@{}}
\hline
\textbf{Scenario} & \textbf{Strict} & \textbf{Gate} & \textbf{Judge} & \textbf{Rule} & \textbf{Safety} & \textbf{Flow} & \textbf{Other}\\
\hline
iCloud & 64 & 62 & 84 & 93 & 80 & 161 & 4 other\\
App permission & 102 & 102 & 104 & 103 & 114 & 115 & 1 keyword\\
Transfer & 90 & 90 & 118 & 97 & 134 & 165 & 2 keyword\\
Calendar & 107 & 50 & 106 & 116 & 133 & 146 & 1 other\\
Driving & 0 & 0 & 4 & 4 & 6 & 6 & 0\\
\hline
\end{tabular*}
\caption{Major sandbox and judge rejection signals. Counts are diagnostic flags and can overlap because a candidate can trigger multiple failure reasons. The recurring state-flow and safety failures explain why sandbox validation is stricter than BestScore@20 or strict text scoring alone.}
\label{tab:app_sandbox_failures}
\end{table*}

\section{Error Analysis and Retrieval Diagnostics}
\label{app:error_recovery}

The main residual failures are contextual boundary errors. In the 832-example joint-field diagnostic, L2 Closed reaches 0.7127 joint accuracy, with 593 correct cases and 239 joint errors. The most common error categories are protected content allowed, unnecessary masking, and confirmation replaced by masking, accounting for 79.5\% of joint errors.

\begin{table}[h]
\centering
\footnotesize
\setlength{\tabcolsep}{3pt}
\begin{tabular*}{\linewidth}{@{}p{.52\linewidth}@{\extracolsep{\fill}}rr@{}}
\hline
\textbf{Joint-field error} & \textbf{Count} & \textbf{Share of errors}\\
\hline
Protected content allowed & 95 & 39.7\%\\
Unnecessary masking & 58 & 24.3\%\\
Confirmation replaced by masking & 37 & 15.5\%\\
Other joint-field errors & 49 & 20.5\%\\
\hline
Total joint errors & 239 & 100.0\%\\
\hline
\end{tabular*}
\caption{Joint-field error categories for L2 Closed. The top three categories account for 79.5\% of all joint errors and mostly reflect contextual boundary decisions.}
\label{tab:app_error_categories}
\end{table}

\begin{table}[h]
\centering
\footnotesize
\setlength{\tabcolsep}{3pt}
\begin{tabular*}{\linewidth}{@{}p{.70\linewidth}@{\extracolsep{\fill}}r@{}}
\hline
\textbf{Retrieval diagnostic} & \textbf{Count}\\
\hline
Wrong-boundary retrieval & 239\\
Weak retrieval evidence & 8\\
Neither & 585\\
\hline
\end{tabular*}
\caption{Retrieval diagnostics for L2 Closed. Wrong-boundary retrieval is a diagnostic category for policy-boundary confusion.}
\label{tab:app_retrieval_diagnostics}
\end{table}

The equality between 239 wrong-boundary retrieval cases and 239 joint errors is numerical. A wrong-boundary retrieval means the retrieved rule is topically related but aligned to the wrong policy boundary. This supports the paper's main interpretation: static rules require reliable applicability judgment and execution gates.

\begin{table*}[t]
\centering
\footnotesize
\setlength{\tabcolsep}{4pt}
\begin{tabular*}{\textwidth}{@{}p{.35\textwidth}@{\extracolsep{\fill}}rrrrp{.18\textwidth}@{}}
\hline
\textbf{Exploratory setting} & \textbf{Recovery} & \textbf{Pres. harm} & \textbf{OOS harm} & \textbf{Net} & \textbf{Use}\\
\hline
Routed skill application & 0 & -- & -- & -- & diagnostic only\\
Sampled targeted skills & 5 & 2 & 5 & -2 & limitation\\
\hline
\end{tabular*}
  \caption{Targeted skill-recovery diagnostics. Targeted skills can recover some cases, but preservation and out-of-scope harms prevent using them as a main full-set performance result.}
\label{tab:app_skill_recovery}
\end{table*}

The skill-recovery results serve as a boundary and future-work signal. Reliable applicability judgment and an execution gate are required before targeted skills can be merged into a general policy system.

\section{Prompt, Mediator, and Judge Templates}
\label{app:prompt_templates}

We include canonical templates as a compact summary of the prompting and mediation interfaces. The templates show the responsibilities, fields, and output contracts needed for benchmark-level reproduction while keeping sensitive release details, author-identifying metadata, and sample-specific expansions out of the paper. Complete sanitized prompts and mediation code belong in the released artifact.

\paragraph{L2 Open policy template.}
The L2 Open baseline receives the raw screenshot and task context; retrieved rules and mediator fields are withheld. Its role is to predict the two canonical decision fields. The instructions frame the task as contextual exposure arbitration: visible sensitive-looking content triggers masking only when policy context requires it, task need triggers confirmation only under the relevant consent boundary, and Ask is reserved for confirmation-sensitive cases. The evaluator derives the final policy label from the two canonical fields.

\begin{quote}
\small
\textbf{Role:} on-device privacy policy checker for whether the original screenshot can be exposed to an untrusted downstream agent under local protection policy.\\
\textbf{Inputs:} app/platform, persona, relationship tag, user task, and screenshot.\\
\textbf{Decision fields:} mask flag and confirmation flag.\\
\textbf{Confirmation trigger:} use confirmation only when recipient identity, authorization boundary, identifier semantics, or visual evidence is unclear.\\
\textbf{Required JSON:} mask flag, confirmation flag, and rationale.
\end{quote}

\paragraph{L2 Closed mediator-locked template.}
The L2 Closed setting inserts retrieval and a deterministic mediator before the local VLM visual check. The retrieved policy rules and mediator decision are rendered into the prompt, and the mediator owns the final policy fields. The VLM checks whether the screenshot contains visual evidence that supports the mediator decision or a clear visual contradiction. When it detects a contradiction, it reports the contradiction while preserving the locked policy fields in the same response.

\begin{quote}
\small
\textbf{Role:} visual-evidence and contradiction checker under mediator-locked policy fields.\\
\textbf{Inputs:} app/platform, persona context, relationship context, user instruction, agent intent, top-k policy rules, mediator-selected rule id, mediator mask flag, mediator confirmation flag, mediator policy decision, and mediator decision basis.\\
\textbf{Locked fields:} selected rule id, mask flag, and confirmation flag must be copied from the mediator values.\\
\textbf{Visual check fields:} visual evidence, visual contradiction flag, and rationale.\\
\textbf{Required JSON:} selected rule id, mask flag, confirmation flag, rationale, visual evidence, and contradiction flag.
\end{quote}

\begin{center}
\begin{minipage}{\linewidth}
\centering
\scriptsize
\setlength{\tabcolsep}{3pt}
\begin{tabular*}{\linewidth}{@{}p{.28\linewidth}@{\extracolsep{\fill}}p{.62\linewidth}@{}}
\hline
\textbf{Field group} & \textbf{Source and overwrite boundary}\\
\hline
Retrieved rules & \texttt{topk\_rules\_block}; retrieval context consumed by the deterministic mediator.\\
Locked policy fields & selected rule id, mask flag, confirmation flag, and derived policy decision; copied by the local VLM and used for scoring.\\
VLM diagnostics & visual evidence and visual contradiction flag; generated by the local VLM but unable to overwrite policy fields.\\
\hline
\end{tabular*}
\captionof{table}{Controlled L2 policy interface. The local VLM adds visual support or contradiction diagnostics while the mediator keeps the policy fields fixed for the benchmark run.}
\label{tab:app_mediator_interface}
\end{minipage}
\end{center}

\paragraph{Deterministic mediator boundary.}
The reported L2 Closed result uses a deterministic mediator instance over structured retrieval-query facets and top-k rule metadata. The mediator maps the retrieved rule context and scenario facets to a selected rule id, a mask flag, a confirmation flag, and a derived policy decision. The implementation uses a deterministic rule-priority and fallback procedure. This design separates variable policy knowledge from a fixed execution interface.

The implementation is treated as experiment-reproduction code. In the paper, we report its interface and reproduction boundary because the claim is about benchmark-level retrieval-augmented policy arbitration. The final L2 Closed setting reaches $0.717{\pm}0.007$ policy accuracy on the 832-sample benchmark.

\paragraph{Rule and skill terminology.}
Rules support policy arbitration in the static L2 mediator. Skills are validated behavior policies used in the self-evolution setting. Feedback may produce candidate constraints, rule-like conditions, or skill patches, and the Experiment 3 claim focuses on behavior-improving skills in controlled scenarios.

\paragraph{Judge and sandbox rubric.}
The judge checks whether a candidate skill specifies the relevant privacy condition, asks for confirmation at the right time, avoids unsafe action defaults, and provides executable state-flow logic. A pass means the candidate satisfies the audit protocol for that controlled scenario.

\section{Implementation Boundary}
\label{app:e2_replay}

The reported experiments evaluate offline policy arbitration, component-level SafeScreenshot construction, and controlled text-skill adaptation. We do not claim end-to-end device latency or deployment readiness. Live deployment would additionally require screenshot capture, model loading, network variability for cloud controls, GUI execution, and user-confirmation latency.

\section{Annotation and Scenario Construction}
\label{app:annotation_instructions}

The benchmark scenarios are reconstructed and sanitized. Scenario construction records the persona abstraction, app/platform, user instruction, candidate agent action, recipient or relationship context, sensitive evidence, and expected policy decision. Labels distinguish whether the raw screenshot can be exposed, whether a SafeScreenshot is required, or whether the system should pause for confirmation. The construction uses author-audited benchmark labels for controlled evaluation, with real user logs, real private chats, external crowdworkers, and human-subject interactions outside the data source.

For Experiment 3, each controlled scenario defines task goals, initial skills, user corrections, distilled constraints, and held-out behavior tests. Candidate skills are evaluated under fixed scoring and sandbox checks. This construction supports controlled adaptation claims over designed feedback signals.

\begin{center}
\begin{minipage}{\linewidth}
\centering
\scriptsize
\setlength{\tabcolsep}{3pt}
\resizebox{\linewidth}{!}{%
\begin{tabular}{p{.28\linewidth}p{.64\linewidth}}
\hline
\textbf{Step} & \textbf{Recorded fields and validation boundary}\\
\hline
UI pattern abstraction & App/platform family, interaction template, layout class, and task surface; patterns approximate common GUI interactions through reconstructed layouts.\\
LLM-assisted scenario drafting & Persona abstraction, task instruction, candidate action, and recipient or relationship context; drafts are prompt generated and sanitized.\\
Policy-label consistency pass & Allow, Mask, Ask, lower-level mask and confirmation fields; inconsistent records are rejected or repaired before scoring.\\
Sensitive-field injection & Typed fake values, visual/textual evidence, localizable-region status, PII/risk category, and benchmark bucket; audit fields stay offline.\\
HTML/UI rendering & Reconstructed layout, render viewport, and screenshot path; rendered screenshots become benchmark and SafeScreenshot inputs.\\
Label correction and audit & Canonical expected-decision table, correction notes, and E3 repair records; known corrections are synchronized before scoring.\\
Skill scenario construction & Task goal, initial skills, correction signals, distilled constraints, and held-out tests; used only for controlled adaptation experiments.\\
\hline
\end{tabular}
}
\captionof{table}{Scenario and label construction protocol. The table records what is controlled during benchmark construction and what remains offline-only audit information.}
\label{tab:app_annotation_protocol}
\end{minipage}
\end{center}

\paragraph{Sanitized instruction excerpt.}
Scenario construction instructions require the project authors to specify: (i) the persona and privacy sensitivity, (ii) the application and recipient boundary, (iii) the candidate agent action, (iv) visible sensitive evidence and whether it is localizable, (v) the expected policy decision, and (vi) a short rationale explaining why the current GUI state should be exposed, masked, or paused for confirmation. For skill-evolution scenarios, the corresponding record also includes user corrections, target constraints, and held-out behavior tests. Because no real users or external annotators are involved, there is no participant compensation or consent process for the dataset construction reported here.

\section{Artifact, Model, and Reproducibility Details}
\label{app:reproducibility}

\begin{center}
\begin{minipage}{\linewidth}
\centering
\footnotesize
\setlength{\tabcolsep}{3pt}
\begin{tabular*}{\linewidth}{@{}p{.20\linewidth}@{\extracolsep{\fill}}p{.72\linewidth}@{}}
\hline
\textbf{Experiment} & \textbf{Inputs, controls, and boundary}\\
\hline
E1 L2 & 832 rows, prompts, policy memory, predictions, and fixed evaluator.\\
E1 L1/L3 & 832 triage rows, 736 OCR-coverage rows, and 558 visible-PII SafeScreenshot rows; component diagnostics only.\\
E2 baseline & 832 rows and prediction files, recomputed against expected labels; routing systems use the route-level mapping in Appendix~\ref{app:baseline_details}.\\
E3 skills & Five scenario specs, correction signals, candidate skills, behavior tests, and recorded seeds.\\
Sandbox & Candidate skills, strict scorer, and LLM-Judge rubric; audit gate for flow-level safety.\\
\hline
\end{tabular*}
\captionof{table}{Experiment-level reproducibility controls. The experiments are inference and text-skill optimization runs; no hidden model parameters are trained.}
\label{tab:app_experiment_settings}
\end{minipage}
\end{center}

Reproduction uses sanitized input tables, prediction files, policy-memory summaries, prompt templates, mediation code, and scoring code. The reported L2 Closed setting uses 40 policy rules (12 Ask, 11 Allow, and 17 Mask), local retrieval over task, app, persona, recipient, and visual evidence, and deterministic mediation. The study uses inference, replay, and text-skill optimization only; no model is trained or fine-tuned.

\begin{center}
\begin{minipage}{\linewidth}
\centering
\footnotesize
\setlength{\tabcolsep}{3pt}
\begin{tabular*}{\linewidth}{@{}p{.20\linewidth}@{\extracolsep{\fill}}p{.72\linewidth}@{}}
\hline
\textbf{Component} & \textbf{Reproducible setting and boundary}\\
\hline
Local VLM & 8B-class local multimodal model; local inference wrapper, inference only.\\
Cloud VLMs & Rule-free cloud controls with raw screenshots, temperature 0.0, and parsed JSON outputs.\\
LLM-Judge & External multimodal judge used only for skill-evolution sandbox checks.\\
Rule retrieval & 40-rule policy memory; local retrieval over task, app, persona, recipient, and visual evidence.\\
Skill evolution & Cold/medium/hot starts, up to 20 iterations, seed-controlled mutation, and validation filters.\\
\hline
\end{tabular*}
\captionof{table}{Model, retrieval, and implementation settings needed to interpret the reported experiments. Provider- or endpoint-specific metadata belongs in the released artifact.}
\label{tab:app_model_settings}
\vspace{0.45em}
\centering
\footnotesize
\setlength{\tabcolsep}{4pt}
\begin{tabular*}{\linewidth}{@{}p{.40\linewidth}@{\extracolsep{\fill}}p{.50\linewidth}@{}}
\hline
\textbf{L1/L3 reproducibility item} & \textbf{Role}\\
\hline
L1 OCR predictions & Text-coverage denominator for the main-table L1 score.\\
Visible-PII L3 input & L3 denominator for visible-PII screenshots.\\
Sensitive items & Sensitive-item extraction output.\\
Strict mask audit & Reported strict SafeScreenshot audit table.\\
Audit manifest & Targeted visual audit pack for manual inspection.\\
\hline
\end{tabular*}
\captionof{table}{Key L1/L3 items for reproducing the SafeScreenshot component audit. Exact filenames belong in the supplementary artifact.}
\label{tab:app_l3_artifacts}
\end{minipage}
\end{center}

\section{Dataset Card, Example Records, and Release Scope}
\label{app:dataset_card}

P-GUI-Evo is a research benchmark for GUI privacy mediation, built from abstracted UI patterns, prompt-drafted scenarios, checked policy labels, sanitized sensitive fields, and reconstructed HTML/UI screenshots. Its intended use is controlled GUI privacy research; private-attribute inference and medical, financial, workplace, or identity decisions are outside scope.

\begin{table*}[t]
\centering
\small
\begin{tabular*}{\textwidth}{@{}p{.18\textwidth}@{\extracolsep{\fill}}p{.50\textwidth}p{.24\textwidth}@{}}
\hline
\textbf{Field group} & \textbf{Examples} & \textbf{Visibility}\\
\hline
Runner input & screenshot, app/platform, persona context, user instruction, agent intent & method input\\
Policy output & Allow, Mask, Ask & prediction/evaluation\\
Construction metadata & UI pattern, prompt template, persona/task/recipient context, HTML/UI layout, render viewport & audit/reproduction\\
Sensitive-field metadata & PII type, fake value map, injection method, localizable status & offline only\\
Audit metadata & risk tag, localizable region, expected label, annotation notes & offline only\\
Release metadata & sample id, bucket, split, sanitized scenario summary & sanitized artifact\\
\hline
\end{tabular*}
\caption{Dataset schema and field visibility. Deployed mediator inputs receive only runner-visible fields.}
\label{tab:app_dataset_schema}
\end{table*}

\begin{table*}[t]
\centering
\footnotesize
\setlength{\tabcolsep}{4pt}
\begin{tabular*}{\textwidth}{@{}p{.20\textwidth}@{\extracolsep{\fill}}p{.72\textwidth}@{}}
\hline
\textbf{Item} & \textbf{Statement}\\
\hline
Intended use & Research benchmark for GUI-agent privacy mediation, policy arbitration, SafeScreenshot construction, and skill-evolution evaluation.\\
Non-use boundary & Not intended for inferring real users' private attributes, identity verification, medical or financial decision making, workplace surveillance, or training systems to expose private visual evidence.\\
Source and construction & Reconstructed GUI interaction patterns: LLM-assisted metadata drafting, policy-label checks, sanitized sensitive-field injection, HTML/UI rendering, and author review.\\
Release artifacts & Sanitized metadata, labels, prompt templates, scoring code, policy-memory summaries, and release-eligible screenshots or masked derivatives after final privacy review. During anonymous review, materials should use an anonymous supplementary artifact or placeholder.\\
Withheld assets & Any image that contains real accounts, contacts, organization identifiers, QR codes, credentials, identity documents, or reversible identifiers should be withheld or replaced by sanitized surrogates.\\
License / terms & The P-GUI-Evo benchmark release is intended under CC BY 4.0 for release-eligible reconstructed metadata, labels, templates, and sanitized visual assets. Third-party models, tools, and API services remain governed by the respective upstream licenses or provider terms, with details recorded in the artifact README/NOTICE file.\\
\hline
\end{tabular*}
\caption{Dataset release and intended-use statement. The release scope is deliberately narrower than the internal experimental workspace.}
\label{tab:app_release_scope}
\end{table*}

\begin{table*}[t]
\centering
\footnotesize
\setlength{\tabcolsep}{3pt}
\begin{tabular*}{\textwidth}{@{}p{.17\textwidth}@{\extracolsep{\fill}}p{.07\textwidth}p{.16\textwidth}p{.40\textwidth}p{.12\textwidth}@{}}
\hline
\textbf{Example} & \textbf{Policy} & \textbf{Context} & \textbf{English gloss} & \textbf{Role}\\
\hline
WeChat medical record transfer & Ask & User A, medical sharing to an external consultant & A reconstructed chat prepares medical details for external consultation; the system should pause and avoid silent exposure. & Ask gate / no masking example\\
QQ bank-card chat & Mask & User C, chat with an online seller involving a financial identifier & A reconstructed QQ conversation contains a bank-card number; the SafeScreenshot masks the financial identifier. & Mask / exposure-denial example\\
Hospital OA travel request & Allow & User A, structured workplace form & A reconstructed workplace form contains employee fields; the example illustrates a task-relevant UI that should remain inspectable. & Allow / no masking example\\
\hline
\end{tabular*}
\caption{Representative reconstructed records corresponding to Figure~\ref{fig:app_examples}. The English gloss explains the privacy boundary while preserving the original benchmark label.}
\label{tab:app_representative_records}
\end{table*}

If screenshots are released, they will receive a final privacy scan for accounts, credentials, QR codes, identifiers, and sensitive records; unreleased visuals will be replaced by sanitized labels, prompts, scripts, schemas, and rendering metadata where permitted.

\begin{figure*}[p]
\centering
\includegraphics[width=.92\textwidth]{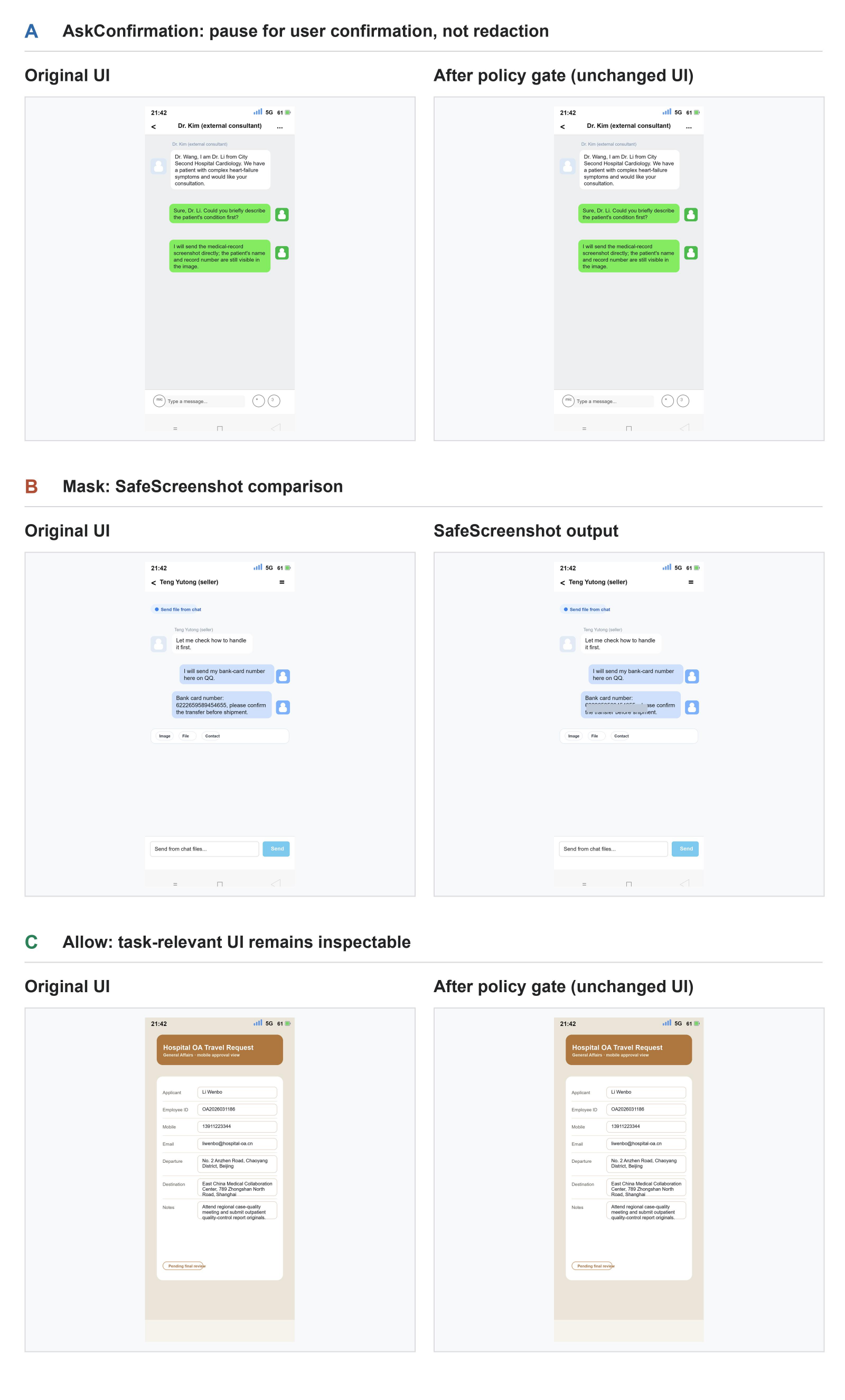}
\caption{Representative reconstructed benchmark examples. Ask and Allow keep the original UI on both sides; Mask shows the before/after SafeScreenshot comparison. All shown visual content is synthetic or reconstructed.}
\label{fig:app_examples}
\end{figure*}

%% file: main.bib
@article{asai2024,
  title = {Self-{{RAG}}: {{Learning}} to {{Retrieve}}, {{Generate}}, and {{Critique}} through {{Self-Reflection}}},
  shorttitle = {Self-{{RAG}}},
  author = {Asai, Akari and Wu, Zeqiu and Wang, Yizhong and Sil, Avi and Hajishirzi, Hannaneh},
  year = 2024,
  month = may,
  journal = {International Conference on Learning Representations},
  volume = {2024},
  pages = {9112--9141},
  langid = {english}
}

@inproceedings{carlini2021,
  title = {Extracting {{Training Data}} from {{Large Language Models}}},
  booktitle = {30th {{USENIX Security Symposium}} ({{USENIX Security}} 21)},
  author = {Carlini, Nicholas and Tram{\`e}r, Florian and Wallace, Eric and Jagielski, Matthew and {Herbert-Voss}, Ariel and Lee, Katherine and Roberts, Adam and Brown, Tom and Song, Dawn and Erlingsson, {\'U}lfar and Oprea, Alina and Raffel, Colin},
  year = 2021,
  pages = {2633--2650},
  langid = {english}
}

@misc{chang2026,
  title = {Know {{Thy Enemy}}: {{Securing LLMs Against Prompt Injection}} via {{Diverse Data Synthesis}} and {{Instruction-Level Chain-of-Thought Learning}}},
  shorttitle = {Know {{Thy Enemy}}},
  author = {Chang, Zhiyuan and Li, Mingyang and Huang, Yuekai and Jiang, Ziyou and Jia, Xiaojun and Xiong, Qian and Wang, Junjie and Li, Zhaoyang and Wang, Qing},
  year = 2026,
  month = apr,
  number = {arXiv:2601.04666},
  eprint = {2601.04666},
  primaryclass = {cs.AI},
  publisher = {arXiv},
  doi = {10.48550/arXiv.2601.04666},
  archiveprefix = {arXiv}
}

@misc{chaudhari2026,
  title = {Assessing {{Privacy Preservation}} and {{Utility}} in {{Online Vision-Language Models}}},
  author = {Chaudhari, Karmesh Siddharam and Zhu, Youxiang and Feng, Amy and Liang, Xiaohui and Zhang, Honggang},
  year = 2026,
  month = apr,
  number = {arXiv:2604.09695},
  eprint = {2604.09695},
  primaryclass = {cs.CV},
  publisher = {arXiv},
  doi = {10.48550/arXiv.2604.09695},
  archiveprefix = {arXiv}
}

@misc{chen2026,
  title = {{{MemPrivacy}}: {{Privacy-Preserving Personalized Memory Management}} for {{Edge-Cloud Agents}}},
  shorttitle = {{{MemPrivacy}}},
  author = {Chen, Yining and Zhao, Jihao and Tang, Bo and Wang, Haofen and Zhang, Yue and Huang, Fei and Xiong, Feiyu and Li, Zhiyu},
  year = 2026,
  month = may,
  number = {arXiv:2605.09530},
  eprint = {2605.09530},
  primaryclass = {cs.CR},
  publisher = {arXiv},
  doi = {10.48550/arXiv.2605.09530},
  archiveprefix = {arXiv}
}

@article{guo2024,
  title = {Connecting {{Large Language Models}} with {{Evolutionary Algorithms Yields Powerful Prompt Optimizers}}},
  author = {Guo, Qingyan and Wang, Rui and Guo, Junliang and Li, Bei and Song, Kaitao and Tan, Xu and Liu, Guoqing and Bian, Jiang and Yang, Yujiu},
  year = 2024,
  month = may,
  journal = {International Conference on Learning Representations},
  volume = {2024},
  pages = {34133--34156},
  langid = {english}
}

@inproceedings{habernal2023,
  title = {Privacy-{{Preserving Natural Language Processing}}},
  booktitle = {Proceedings of the 17th {{Conference}} of the {{European Chapter}} of the {{Association}} for {{Computational Linguistics}}: {{Tutorial Abstracts}}},
  author = {Habernal, Ivan and Mireshghallah, Fatemehsadat and Thaine, Patricia and Ghanavati, Sepideh and Feyisetan, Oluwaseyi},
  editor = {Zanzotto, Fabio Massimo and Pradhan, Sameer},
  year = 2023,
  month = may,
  pages = {27--30},
  publisher = {Association for Computational Linguistics},
  address = {Dubrovnik, Croatia},
  doi = {10.18653/v1/2023.eacl-tutorials.6}
}

@inproceedings{hong2024,
  title = {{{CogAgent}}: {{A Visual Language Model}} for {{GUI Agents}}},
  shorttitle = {{{CogAgent}}},
  booktitle = {Proceedings of the {{IEEE}}/{{CVF Conference}} on {{Computer Vision}} and {{Pattern Recognition}}},
  author = {Hong, Wenyi and Wang, Weihan and Lv, Qingsong and Xu, Jiazheng and Yu, Wenmeng and Ji, Junhui and Wang, Yan and Wang, Zihan and Dong, Yuxiao and Ding, Ming and Tang, Jie},
  year = 2024,
  pages = {14281--14290},
  langid = {english}
}

@misc{hu2024,
  title = {{{MiniCPM}}: {{Unveiling}} the {{Potential}} of {{Small Language Models}} with {{Scalable Training Strategies}}},
  shorttitle = {{{MiniCPM}}},
  author = {Hu, Shengding and Tu, Yuge and Han, Xu and He, Chaoqun and Cui, Ganqu and Long, Xiang and Zheng, Zhi and Fang, Yewei and Huang, Yuxiang and Zhao, Weilin and Zhang, Xinrong and Thai, Zheng Leng and Zhang, Kaihuo and Wang, Chongyi and Yao, Yuan and Zhao, Chenyang and Zhou, Jie and Cai, Jie and Zhai, Zhongwu and Ding, Ning and Jia, Chao and Zeng, Guoyang and Li, Dahai and Liu, Zhiyuan and Sun, Maosong},
  year = 2024,
  publisher = {arXiv},
  doi = {10.48550/ARXIV.2404.06395}
}

@misc{jiang2026,
  title = {{{TreeCUA}}: {{Efficiently Scaling GUI Automation}} with {{Tree-Structured Verifiable Evolution}}},
  shorttitle = {{{TreeCUA}}},
  author = {Jiang, Deyang and Huang, Jing and Zhao, Xuanle and Chen, Lei and Zheng, Liming and Liu, Fanfan and Qiu, Haibo and Shi, Peng and Zeng, Zhixiong},
  year = 2026,
  month = feb,
  journal = {arXiv.org},
  howpublished = {https://arxiv.org/abs/2602.09662v1},
  langid = {english}
}

@article{kim2023,
  title = {{{ProPILE}}: {{Probing Privacy Leakage}} in {{Large Language Models}}},
  shorttitle = {{{ProPILE}}},
  author = {Kim, Siwon and Yun, Sangdoo and Lee, Hwaran and Gubri, Martin and Yoon, Sungroh and Oh, Seong Joon},
  year = 2023,
  month = dec,
  journal = {Advances in Neural Information Processing Systems},
  volume = {36},
  pages = {20750--20762},
  langid = {english}
}

@inproceedings{lewis2020,
  title = {Retrieval-{{Augmented Generation}} for {{Knowledge-Intensive NLP Tasks}}},
  booktitle = {Advances in {{Neural Information Processing Systems}}},
  author = {Lewis, Patrick and Perez, Ethan and Piktus, Aleksandra and Petroni, Fabio and Karpukhin, Vladimir and Goyal, Naman and K{\"u}ttler, Heinrich and Lewis, Mike and Yih, Wen-tau and Rockt{\"a}schel, Tim and Riedel, Sebastian and Kiela, Douwe},
  year = 2020,
  volume = {33},
  pages = {9459--9474},
  publisher = {Curran Associates, Inc.}
}

@misc{lin2025,
  title = {Computer-{{Use Agents}} as {{Judges}} for {{Generative User Interface}}},
  author = {Lin, Kevin Qinghong and Hu, Siyuan and Li, Linjie and Yang, Zhengyuan and Wang, Lijuan and Torr, Philip and Shou, Mike Zheng},
  year = 2025,
  publisher = {arXiv},
  doi = {10.48550/ARXIV.2511.15567}
}

@misc{liu2024,
  title = {{{AutoGLM}}: {{Autonomous Foundation Agents}} for {{GUIs}}},
  shorttitle = {{{AutoGLM}}},
  author = {Liu, Xiao and Qin, Bo and Liang, Dongzhu and Dong, Guang and Lai, Hanyu and Zhang, Hanchen and Zhao, Hanlin and Iong, Iat Long and Sun, Jiadai and Wang, Jiaqi and Gao, Junjie and Shan, Junjun and Liu, Kangning and Zhang, Shudan and Yao, Shuntian and Cheng, Siyi and Yao, Wentao and Zhao, Wenyi and Liu, Xinghan and Liu, Xinyi and Chen, Xinying and Yang, Xinyue and Yang, Yang and Xu, Yifan and Yang, Yu and Wang, Yujia and Xu, Yulin and Qi, Zehan and Dong, Yuxiao and Tang, Jie},
  year = 2024,
  month = oct,
  journal = {arXiv.org},
  howpublished = {https://arxiv.org/abs/2411.00820v1},
  langid = {english}
}

@article{lu2024,
  title = {{{WildVision}}: {{Evaluating Vision-Language Models}} in the {{Wild}} with {{Human Preferences}}},
  shorttitle = {{{WildVision}}},
  author = {Lu, Yujie and Jiang, Dongfu and Chen, Wenhu and Wang, William Yang and Choi, Yejin and Lin, Bill Yuchen},
  year = 2024,
  month = dec,
  journal = {Advances in Neural Information Processing Systems},
  volume = {37},
  pages = {48224--48255},
  doi = {10.52202/079017-1528},
  langid = {english}
}

@misc{lu2024b,
  title = {{{OmniParser}} for {{Pure Vision Based GUI Agent}}},
  author = {Lu, Yadong and Yang, Jianwei and Shen, Yelong and Awadallah, Ahmed},
  year = 2024,
  month = aug,
  number = {arXiv:2408.00203},
  eprint = {2408.00203},
  primaryclass = {cs.CV},
  publisher = {arXiv},
  doi = {10.48550/arXiv.2408.00203},
  archiveprefix = {arXiv}
}

@misc{ma2026,
  title = {{{SkillClaw}}: {{Let Skills Evolve Collectively}} with {{Agentic Evolver}}},
  shorttitle = {{{SkillClaw}}},
  author = {Ma, Ziyu and Yang, Shidong and Ji, Yuxiang and Wang, Xucong and Wang, Yong and Hu, Yiming and Huang, Tongwen and Chu, Xiangxiang},
  year = 2026,
  month = apr,
  number = {arXiv:2604.08377},
  eprint = {2604.08377},
  primaryclass = {cs.AI},
  publisher = {arXiv},
  doi = {10.48550/arXiv.2604.08377},
  archiveprefix = {arXiv}
}

@article{nissenbaum2004a,
  title = {Privacy as {{Contextual Integrity}}},
  author = {Nissenbaum, Helen},
  year = 2004,
  month = feb,
  journal = {Washington Law Review},
  volume = {79},
  number = {1},
  pages = {119}
}

@misc{packer2023,
  title = {{{MemGPT}}: {{Towards LLMs}} as {{Operating Systems}}},
  shorttitle = {{{MemGPT}}},
  author = {Packer, Charles and Wooders, Sarah and Lin, Kevin and Fang, Vivian and Patil, Shishir G. and Stoica, Ion and Gonzalez, Joseph E.},
  year = 2023,
  publisher = {arXiv},
  doi = {10.48550/ARXIV.2310.08560}
}

@article{qi2025,
  title = {{{WebRL}}: {{Training LLM Web Agents}} via {{Self-Evolving Online Curriculum Reinforcement Learning}}},
  shorttitle = {{{WebRL}}},
  author = {Qi, Zehan and Liu, Xiao and Iong, Iat Long and Lai, Hanyu and Sun, Xueqiao and Sun, Jiadai and Yang, Xinyue and Yang, Yu and Yao, Shuntian and Xu, Wei and Tang, Jie and Dong, Yuxiao},
  year = 2025,
  month = may,
  journal = {International Conference on Learning Representations},
  volume = {2025},
  pages = {79791--79821},
  langid = {english}
}

@article{rawles2023,
  title = {{{AndroidInTheWild}}: {{A Large-Scale Dataset For Android Device Control}}},
  shorttitle = {{{AndroidInTheWild}}},
  author = {Rawles, Christopher and Li, Alice and Rodriguez, Daniel and Riva, Oriana and Lillicrap, Timothy},
  year = 2023,
  month = dec,
  journal = {Advances in Neural Information Processing Systems},
  volume = {36},
  pages = {59708--59728},
  langid = {english}
}

@misc{salemi2023,
  title = {{{LaMP}}: {{When Large Language Models Meet Personalization}}},
  shorttitle = {{{LaMP}}},
  author = {Salemi, Alireza and Mysore, Sheshera and Bendersky, Michael and Zamani, Hamed},
  year = 2023,
  publisher = {arXiv},
  doi = {10.48550/ARXIV.2304.11406}
}

@article{shinn2023,
  title = {Reflexion: Language Agents with Verbal Reinforcement Learning},
  shorttitle = {Reflexion},
  author = {Shinn, Noah and Cassano, Federico and Gopinath, Ashwin and Narasimhan, Karthik and Yao, Shunyu},
  year = 2023,
  month = dec,
  journal = {Advances in Neural Information Processing Systems},
  volume = {36},
  pages = {8634--8652},
  langid = {english}
}

@inproceedings{siyan2025,
  title = {{{PAPILLON}}: {{Privacy Preservation}} from {{Internet-based}} and {{Local Language Model Ensembles}}},
  shorttitle = {{{PAPILLON}}},
  booktitle = {Proceedings of the 2025 {{Conference}} of the {{Nations}} of the {{Americas Chapter}} of the {{Association}} for {{Computational Linguistics}}: {{Human Language Technologies}} ({{Volume}} 1: {{Long Papers}})},
  author = {Siyan, Li and Raghuram, Vethavikashini Chithrra and Khattab, Omar and Hirschberg, Julia and Yu, Zhou},
  editor = {Chiruzzo, Luis and Ritter, Alan and Wang, Lu},
  year = 2025,
  month = apr,
  pages = {3371--3390},
  publisher = {Association for Computational Linguistics},
  address = {Albuquerque, New Mexico},
  doi = {10.18653/v1/2025.naacl-long.173}
}

@article{tomekce2024,
  title = {Private {{Attribute Inference}} from {{Images}} with {{Vision-Language Models}}},
  author = {T{\"o}mek{\c c}e, Batuhan and Vero, Mark and Staab, Robin and Vechev, Martin},
  year = 2024,
  month = dec,
  journal = {Advances in Neural Information Processing Systems},
  volume = {37},
  pages = {103619--103651},
  doi = {10.52202/079017-3291},
  langid = {english}
}

@misc{wang2024,
  title = {Mobile-{{Agent}}: {{Autonomous Multi-Modal Mobile Device Agent}} with {{Visual Perception}}},
  shorttitle = {Mobile-{{Agent}}},
  author = {Wang, Junyang and Xu, Haiyang and Ye, Jiabo and Yan, Ming and Shen, Weizhou and Zhang, Ji and Huang, Fei and Sang, Jitao},
  year = 2024,
  publisher = {arXiv},
  doi = {10.48550/ARXIV.2401.16158}
}

@misc{wang2025,
  title = {{{GUI Agents}} with {{Foundation Models}}: {{A Comprehensive Survey}}},
  shorttitle = {{{GUI Agents}} with {{Foundation Models}}},
  author = {Wang, Shuai and Liu, Weiwen and Chen, Jingxuan and Zhou, Yuqi and Gan, Weinan and Zeng, Xingshan and Che, Yuhan and Yu, Shuai and Hao, Xinlong and Shao, Kun and Wang, Bin and Wu, Chuhan and Wang, Yasheng and Tang, Ruiming and Hao, Jianye},
  year = 2025,
  month = feb,
  number = {arXiv:2411.04890},
  eprint = {2411.04890},
  primaryclass = {cs.AI},
  publisher = {arXiv},
  doi = {10.48550/arXiv.2411.04890},
  archiveprefix = {arXiv}
}

@misc{wang2025a,
  title = {Privacy in {{Action}}: {{Towards Realistic Privacy Mitigation}} and {{Evaluation}} for {{LLM-Powered Agents}}},
  shorttitle = {Privacy in {{Action}}},
  author = {Wang, Shouju and Yu, Fenglin and Liu, Xirui and Qin, Xiaoting and Zhang, Jue and Lin, Qingwei and Zhang, Dongmei and Rajmohan, Saravan},
  year = 2025,
  month = sep,
  number = {arXiv:2509.17488},
  eprint = {2509.17488},
  primaryclass = {cs.CR},
  publisher = {arXiv},
  doi = {10.48550/arXiv.2509.17488},
  archiveprefix = {arXiv}
}

@inproceedings{wang2025b,
  title = {{{PIG}}: {{Privacy Jailbreak Attack}} on {{LLMs}} via {{Gradient-based Iterative In-Context Optimization}}},
  shorttitle = {{{PIG}}},
  booktitle = {Proceedings of the 63rd {{Annual Meeting}} of the {{Association}} for {{Computational Linguistics}} ({{Volume}} 1: {{Long Papers}})},
  author = {Wang, Yidan and Cao, Yanan and Ren, Yubing and Fang, Fang and Lin, Zheng and Fang, Binxing},
  editor = {Che, Wanxiang and Nabende, Joyce and Shutova, Ekaterina and Pilehvar, Mohammad Taher},
  year = 2025,
  month = jul,
  pages = {9645--9660},
  publisher = {Association for Computational Linguistics},
  address = {Vienna, Austria},
  doi = {10.18653/v1/2025.acl-long.475}
}

@misc{wang2026,
  title = {{{VisualLeakBench}}: {{Auditing}} the {{Fragility}} of {{Large Vision-Language Models}} against {{PII Leakage}} and {{Social Engineering}}},
  shorttitle = {{{VisualLeakBench}}},
  author = {Wang, Youting and Tang, Yuan and Qian, Yitian and Zhao, Chen},
  year = 2026,
  month = mar,
  number = {arXiv:2603.13385},
  eprint = {2603.13385},
  primaryclass = {cs.CV},
  publisher = {arXiv},
  doi = {10.48550/arXiv.2603.13385},
  archiveprefix = {arXiv}
}

@misc{wang2026a,
  title = {{{GUIGuard-Bench}}: {{Toward}} a {{General Evaluation}} for {{Privacy-Preserving GUI Agents}}},
  shorttitle = {{{GUIGuard-Bench}}},
  author = {Wang, Yanxi and Zhang, Zhiling and Zhou, Wenbo and Zhang, Weiming and Zhang, Jie and Zhu, Qiannan and Shi, Yu and Zheng, Shuxin and He, Jiyan},
  year = 2026,
  publisher = {arXiv},
  doi = {10.48550/ARXIV.2601.18842}
}

@misc{xia2026,
  title = {{{SkillRL}}: {{Evolving Agents}} via {{Recursive Skill-Augmented Reinforcement Learning}}},
  shorttitle = {{{SkillRL}}},
  author = {Xia, Peng and Chen, Jianwen and Wang, Hanyang and Liu, Jiaqi and Zeng, Kaide and Wang, Yu and Han, Siwei and Zhou, Yiyang and Zhao, Xujiang and Chen, Haifeng and Zheng, Zeyu and Xie, Cihang and Yao, Huaxiu},
  year = 2026,
  publisher = {arXiv},
  doi = {10.48550/ARXIV.2602.08234}
}

@inproceedings{xu2025,
  title = {{{AndroidLab}}: {{Training}} and {{Systematic Benchmarking}} of {{Android Autonomous Agents}}},
  shorttitle = {{{AndroidLab}}},
  booktitle = {Proceedings of the 63rd {{Annual Meeting}} of the {{Association}} for {{Computational Linguistics}} ({{Volume}} 1: {{Long Papers}})},
  author = {Xu, Yifan and Liu, Xiao and Sun, Xueqiao and Cheng, Siyi and Yu, Hao and Lai, Hanyu and Zhang, Shudan and Zhang, Dan and Tang, Jie and Dong, Yuxiao},
  editor = {Che, Wanxiang and Nabende, Joyce and Shutova, Ekaterina and Pilehvar, Mohammad Taher},
  year = 2025,
  month = jul,
  pages = {2144--2166},
  publisher = {Association for Computational Linguistics},
  address = {Vienna, Austria},
  doi = {10.18653/v1/2025.acl-long.107}
}

@misc{yan2025,
  title = {Step-{{GUI Technical Report}}},
  author = {Yan, Haolong and Wang, Jia and Huang, Xin and Shen, Yeqing and Meng, Ziyang and Fan, Zhimin and Tan, Kaijun and Gao, Jin and Shi, Lieyu and Yang, Mi and Yang, Shiliang and Wang, Zhirui and Li, Brian and An, Kang and Li, Chenyang and Lei, Lei and Duan, Mengmeng and Liang, Danxun and Liu, Guodong and Cheng, Hang and Wu, Hao and Dong, Jie and Huang, Junhao and Chen, Mei and Yu, Renjie and Li, Shunshan and Zhou, Xu and Dai, Yiting and Deng, Yineng and Liang, Yingdan and Chen, Zelin and Sun, Wen and Yan, Chengxu and Xu, Chunqin and Li, Dong and Xiao, Fengqiong and Fan, Guanghao and Li, Guopeng and Peng, Guozhen and Li, Hongbing and Li, Hang and Chen, Hongming and Xie, Jingjing and Li, Jianyong and Zhang, Jingyang and Ren, Jiaju and Yuan, Jiayu and Yin, Jianpeng and Cao, Kai and Zhao, Liang and Tan, Liguo and Shi, Liying and Ren, Mengqiang and Xu, Min and Liu, Manjiao and Luo, Mao and Wan, Mingxin and Wang, Na and Wu, Nan and Wang, Ning and Ma, Peiyao and Zhang, Qingzhou and Wang, Qiao and Zeng, Qinlin and Gao, Qiong and Li, Qiongyao and Zhong, Shangwu and Gao, Shuli and Liu, Shaofan and Gao, Shisi and Luo, Shuang and Liu, Xingbin and Liu, Xiaojia and Hou, Xiaojie and Liu, Xin and Feng, Xuanti and Cai, Xuedan and Wen, Xuan and Zhu, Xianwei and Liang, Xin and Liu, Xin and Zhou, Xin and Sui, Yifan and Zhao, Yingxiu and Shi, Yukang and Xu, Yunfang and Zeng, Yuqing and Zhang, Yixun and Weng, Zejia and Yan, Zhonghao and Huang, Zhiguo and Wang, Zhuoyu and Yan, Zihan and Ge, Zheng and Li, Jing and Zhu, Yibo and Jiao, Binxing and Zhang, Xiangyu and Jiang, Daxin},
  year = 2025,
  month = dec,
  number = {arXiv:2512.15431},
  eprint = {2512.15431},
  primaryclass = {cs.CV},
  publisher = {arXiv},
  doi = {10.48550/arXiv.2512.15431},
  archiveprefix = {arXiv}
}

@misc{yang2023,
  title = {Set-of-{{Mark Prompting Unleashes Extraordinary Visual Grounding}} in {{GPT-4V}}},
  author = {Yang, Jianwei and Zhang, Hao and Li, Feng and Zou, Xueyan and Li, Chunyuan and Gao, Jianfeng},
  year = 2023,
  month = oct,
  journal = {arXiv.org},
  howpublished = {https://arxiv.org/abs/2310.11441v2},
  langid = {english}
}

@article{yang2024,
  title = {Large {{Language Models}} as {{Optimizers}}},
  author = {Yang, Chengrun and Wang, Xuezhi and Lu, Yifeng and Liu, Hanxiao and Le, Quoc V. and Zhou, Denny and Chen, Xinyun},
  year = 2024,
  month = may,
  journal = {International Conference on Learning Representations},
  volume = {2024},
  pages = {12028--12068},
  langid = {english}
}

@misc{yao2024,
  title = {{{MiniCPM-V}}: {{A GPT-4V Level MLLM}} on {{Your Phone}}},
  shorttitle = {{{MiniCPM-V}}},
  author = {Yao, Yuan and Yu, Tianyu and Zhang, Ao and Wang, Chongyi and Cui, Junbo and Zhu, Hongji and Cai, Tianchi and Li, Haoyu and Zhao, Weilin and He, Zhihui and Chen, Qianyu and Zhou, Huarong and Zou, Zhensheng and Zhang, Haoye and Hu, Shengding and Zheng, Zhi and Zhou, Jie and Cai, Jie and Han, Xu and Zeng, Guoyang and Li, Dahai and Liu, Zhiyuan and Sun, Maosong},
  year = 2024,
  publisher = {arXiv},
  doi = {10.48550/ARXIV.2408.01800}
}

@misc{you2024,
  title = {Ferret-{{UI}}: {{Grounded Mobile UI Understanding}} with {{Multimodal LLMs}}},
  shorttitle = {Ferret-{{UI}}},
  author = {You, Keen and Zhang, Haotian and Schoop, Eldon and Weers, Floris and Swearngin, Amanda and Nichols, Jeffrey and Yang, Yinfei and Gan, Zhe},
  year = 2024,
  month = apr,
  number = {arXiv:2404.05719},
  eprint = {2404.05719},
  primaryclass = {cs.CV},
  publisher = {arXiv},
  doi = {10.48550/arXiv.2404.05719},
  archiveprefix = {arXiv}
}

@misc{yuksekgonul2024a,
  title = {{{TextGrad}}: {{Automatic}} "{{Differentiation}}" via {{Text}}},
  shorttitle = {{{TextGrad}}},
  author = {Yuksekgonul, Mert and Bianchi, Federico and Boen, Joseph and Liu, Sheng and Huang, Zhi and Guestrin, Carlos and Zou, James},
  year = 2024,
  month = jun,
  number = {arXiv:2406.07496},
  eprint = {2406.07496},
  primaryclass = {cs.CL},
  publisher = {arXiv},
  doi = {10.48550/arXiv.2406.07496},
  archiveprefix = {arXiv}
}

@article{zelikman2022,
  title = {{{STaR}}: {{Bootstrapping Reasoning With Reasoning}}},
  shorttitle = {{{STaR}}},
  author = {Zelikman, Eric and Wu, Yuhuai and Mu, Jesse and Goodman, Noah},
  year = 2022,
  month = dec,
  journal = {Advances in Neural Information Processing Systems},
  volume = {35},
  pages = {15476--15488},
  langid = {english}
}

@inproceedings{zhang2025,
  title = {{{AppAgent}}: {{Multimodal Agents}} as {{Smartphone Users}}},
  shorttitle = {{{AppAgent}}},
  booktitle = {Proceedings of the 2025 {{CHI Conference}} on {{Human Factors}} in {{Computing Systems}}},
  author = {Zhang, Chi and Yang, Zhao and Liu, Jiaxuan and Li, Yanda and Han, Yucheng and Chen, Xin and Huang, Zebiao and Fu, Bin and Yu, Gang},
  year = 2025,
  month = apr,
  series = {{{CHI}} '25},
  pages = {1--20},
  publisher = {Association for Computing Machinery},
  address = {New York, NY, USA},
  doi = {10.1145/3706598.3713600}
}

@misc{zhang2026,
  title = {Multi-{{PA}}: {{A Multi-perspective Benchmark}} on {{Privacy Assessment}} for {{Large Vision-Language Models}}},
  shorttitle = {Multi-{{PA}}},
  author = {Zhang, Jie and Cao, Xiangkui and Han, Zhouyu and Shan, Shiguang and Chen, Xilin},
  year = 2026,
  month = mar,
  number = {arXiv:2412.19496},
  eprint = {2412.19496},
  primaryclass = {cs.CR},
  publisher = {arXiv},
  doi = {10.48550/arXiv.2412.19496},
  archiveprefix = {arXiv}
}

@misc{zhao2026,
  title = {Anonymization-{{Enhanced Privacy Protection}} for {{Mobile GUI Agents}}: {{Available}} but {{Invisible}}},
  shorttitle = {Anonymization-{{Enhanced Privacy Protection}} for {{Mobile GUI Agents}}},
  author = {Zhao, Lepeng and Zou, Zhenhua and Li, Shuo and Liu, Zhuotao},
  year = 2026,
  month = feb,
  journal = {arXiv.org},
  howpublished = {https://arxiv.org/abs/2602.10139v3},
  langid = {english}
}

@article{zheng2023,
  title = {Judging {{LLM-as-a-Judge}} with {{MT-Bench}} and {{Chatbot Arena}}},
  author = {Zheng, Lianmin and Chiang, Wei-Lin and Sheng, Ying and Zhuang, Siyuan and Wu, Zhanghao and Zhuang, Yonghao and Lin, Zi and Li, Zhuohan and Li, Dacheng and Xing, Eric and Zhang, Hao and Gonzalez, Joseph and Stoica, Ion},
  year = 2023,
  month = dec,
  journal = {Advances in Neural Information Processing Systems},
  volume = {36},
  pages = {46595--46623},
  langid = {english}
}

@misc{zheng2024,
  title = {{{GPT-4V}}(Ision) Is a {{Generalist Web Agent}}, If {{Grounded}}},
  author = {Zheng, Boyuan and Gou, Boyu and Kil, Jihyung and Sun, Huan and Su, Yu},
  year = 2024,
  month = mar,
  number = {arXiv:2401.01614},
  eprint = {2401.01614},
  primaryclass = {cs.IR},
  publisher = {arXiv},
  doi = {10.48550/arXiv.2401.01614},
  archiveprefix = {arXiv}
}

@misc{zhou2026,
  title = {Memento-{{Skills}}: {{Let Agents Design Agents}}},
  shorttitle = {Memento-{{Skills}}},
  author = {Zhou, Huichi and Guo, Siyuan and Liu, Anjie and Yu, Zhongwei and Gong, Ziqin and Zhao, Bowen and Chen, Zhixun and Zhang, Menglong and Chen, Yihang and Li, Jinsong and Yang, Runyu and Liu, Qiangbin and Yu, Xinlei and Zhou, Jianmin and Wang, Na and Sun, Chunyang and Wang, Jun},
  year = 2026,
  month = mar,
  journal = {arXiv.org},
  howpublished = {https://arxiv.org/abs/2603.18743v1},
  langid = {english}
}

@inproceedings{zhu2024,
  title = {Exploiting {{Privacy Preserving Prompt Techniques}} for {{Online Large Language Model Usage}}},
  author = {Zhu, Youxiang and Gao, Ning and Liang, Xiaohui and Zhang, Honggang},
  year = 2024,
  month = dec,
  booktitle = {GLOBECOM 2024 - 2024 IEEE Global Communications Conference},
  volume = {2024},
  pages = {4304--4309},
  doi = {10.1109/globecom52923.2024.10901426},
  pmcid = {PMC12439101},
  pmid = {40963721}
}
